\documentclass[prd,aps,showpacs,preprintnumbers,amssymb]{revtex4}
\usepackage{graphicx}
\usepackage{epsf}
\usepackage{amsmath}
\usepackage{epstopdf}
\usepackage{bm}
\def\CL{\,\,\, \longrightarrow^{^{^{\hskip -0.55cm C.L.}}} \,\,\,}

\begin{document}

\title{%
\hfill{\normalsize\vbox{%
 }}\\
{Chiral Nonet Mixing in
$\pi K$ Scattering}}

\author{Amir H. Fariborz
$^{\it \bf a}$~\footnote[1]{Email:
 fariboa@sunyit.edu}}

\author{Esmaiel Pourjafarabadi
$^{\it \bf b}$~\footnote[2]{Email:
   esmaiel.pourjafar@shirazu.ac.ir}}

\author{Soodeh Zarepour
	$^{\it \bf b}$~\footnote[5]{Email:
		   soodehzarepour@shirazu.ac.ir         }}

\author{S. Mohammad Zebarjad
$^{\it \bf b}$~\footnote[5]{Email:
   zebarjad@susc.ac.ir        }}

\affiliation{$^ {\bf \it a}$ Department of Mathematics and Physics,
State University of New York, Polytechnic Institute, Utica, NY 13502, USA,}

\affiliation{$^ {\bf \it b}$ Physics Department and Biruni Observatory,  Shiraz University, Shiraz 71454, Iran, }

\date{\today}

\begin{abstract}

The underlying mixing of quark components of scalar mesons is probed in $\pi K$ scattering within a generalized linear sigma model that contains two scalar meson nonets and two pseudoscalar meson nonets (a quark-antiquark  and a four-quark).   In the leading order of this model,  all free parameters have been previously fixed using the mass spectra and several low-energy parameters known from experiment and consistent predictions have been made.  As other predictions of the model,  in the present work the isospins $I$= 1/2, 3/2 and $J$=0 projection of  $\pi K$ scattering amplitude (as well as phase shifts) are computed and compared with experiment.   In the $I$=1/2 channel, it is shown that within the uncertainties of the model parameters a good agreement with experimental data up to an energy of about 1 GeV is obtained, whereas in the $I$=3/2 channel there is a better agreement with experiment which extends to about 1.4 GeV.   The effect of final state interactions of $\pi K$ in the $I$=1/2 channel is approximated by the K-matrix method and the poles of the unitarized scattering amplitude are found.    It is shown that the model predicts a  light and broad kappa  resonance with a mass and decay width of 670-770 MeV and 640-750 MeV consistent with other prior works.    Moreover, the scattering lengths in the $I$=1/2, 3/2 are also computed and shown to qualitatively agree  with experiment.   The overall predictions presented here further support previous findings that the scalar mesons below and above 1 GeV have substantial underlying mixings and that those below 1 GeV have dominant four-quark substructures  while those  above 1 GeV are closer to conventional $P$-wave quark-antiquark states. 

\end{abstract}


\maketitle
\section{introduction}

Scalar mesons are progressively gaining attention for their important roles in low-energy QCD.     They induce spontaneous chiral symmetry breaking and therefore probe the QCD vacuum,  and also they appear as intermediate states in Goldstone boson interactions away from threshold (such as in $\pi\pi$, $\pi K$ and $\pi\eta$ scatterings) in a range of energy that is too low for a perturbative QCD study and too high in the context of chiral perturbation theory.    Moreover,  the scalar mesons have connections to important issues in QCD such as violation of isospin,  diquarks and glueballs.    A general discussion of the experimental  situation on light scalars is given in Ref. \cite{pdg}.

Understanding the physical properties of scalars (particularly their quark substructure) is known to be quite complicated.  Many investigators have tackled different challenging aspects of scalar mesons from different angles (see Refs. \cite{Jaf}-\cite{close}).
The lowest-lying  scalars (below 1 GeV) have a mass spectrum that is much lighter than expected and is also inverted.    This immediately rules out a naive simple quark-antiquark substructure for these states; a picture that is known to work reasonably well for other spins such as vectors and pseudoscalars.   A fundamental framework for understanding the lowest-lying scalars was first proposed by Jaffe in MIT bag model \cite{Jaf}, in which a four-quark (i.e. two quarks and two antiquarks) substructure is considered.    This picture provides an explanation for the light and inverted mass spectrum of the scalars below 1 GeV.  Scalars above 1 GeV are expected to be closer to  quark-antiquark states, but they too have some peculiar properties that make their identification with pure quark-antiquark states rather questionable.   Then it is natural to wonder whether the complexities of the scalars below and above 1 GeV have some underlying connections and whether a global study of these states that includes possible underlying mixings can be useful.   This global treatment defines the philosophy of the framework developed in \cite{global} (and references therein) upon which the present work is built.

In the global study of Ref. \cite{global},  a generalized linear sigma model which is formulated in terms of two scalar nonets and two pseudoscalar nonets (a two- and a four-quark nonet) and the underlying mixings among the scalars and among pseudoscalars is studied.
It is shown that a simple picture for scalar states
below 2 GeV seems to emerge which automatically
leads to light scalars that are dominantly of  two quark-two antiquark
 nature and light
conventional pseudoscalars that are, as expected
from established phenomenology,  dominantly
of quark-antiquark nature.   The free parameters of the leading order of the model are completely determined by fitting to various low-energy data.   After these parameters are determined several  predictions of the model are studied in follow up works, including the prediction of $\pi\pi$ scattering amplitude in \cite{mixing_pipi} which, after unitarizing the amplitude using the  K-matrix method,  gives a reasonable agreement with data up to around 1 GeV.    In the work of \cite{mixing_pipi}, the poles of the K-matrix unitarized amplitude are determined and used to calculate the physical masses and decay widths of the isosinglet scalar mesons.       The first pole corresponded to the light and broad sigma meson with mass and decay width:
\begin{eqnarray}
m \left[f_1\right] &=& 477 \pm 8 \, {\rm MeV}, \nonumber \\
\Gamma\left[f_1\right] &=& 398 \pm 107 \, {\rm MeV}.
\end{eqnarray}
This prediction is consistent with the work of Caprini, Colangelo and  Leutwyler \cite{06_CCL} based on the  Roy equation, for isoscalar S-wave,  in which they find:
\begin{eqnarray}
M_\sigma &=& 441^{+16}_{-8} \, {\rm MeV}, \nonumber \\
\Gamma_\sigma &=& 544^{+16}_{-8}  \, {\rm MeV}.
\end{eqnarray}

The second pole found in \cite{mixing_pipi} resembled the $f_0(980)$:
\begin{eqnarray}
m\left[f_2\right] &=& 1100 \pm 10 \, {\rm MeV}, \nonumber \\
\Gamma\left[f_2\right] &=& 199 \pm 15 \, {\rm MeV}.
\end{eqnarray}
The K-matrix seems to capture the effect of final-state interactions of pions,  which manifest themselves as shifts in the mass and decay width of sigma meson from their Lagrangian values to their physical values (given above), and can be a useful tool for investigating  light and broad scalar mesons.    The main advantages of K-matrix are the fact that it enforces exact unitarity,  and that it does not introduce any additional parameters,  and as such,  acts as a useful mapping of the model predictions to the appropriate experimental data.
The next natural state to investigate is the kappa meson which is probed in $I=1/2$, $J=0$, $\pi K$ scattering amplitude, and will be studied in the present work.     Since all the parameters of the generalized linear sigma model of Ref. \cite{global} have been previously fixed (in its leading order),  the analysis of the $\pi K$ scattering in the present investigation will be a prediction and will further test the model and the underlying mixing patterns predicted in \cite{global}.

The $\pi K$ scattering has been an active topic of both expermental \cite{DIRAC_14,Aston} as well as theoretical \cite{RChPT_bernard_14}-\cite{Lohse_90} investigation in low-energy QCD.  The theoretical analyses  include the chiral perturbation theory \cite{RChPT_bernard_14}-\cite{ChPT_LO_NLO}, the lattice QCD \cite{LQCD_Lang_12}-\cite{NPLQCD_06},  the Roy-Steiner representation \cite{Roy_Steiner_06}-\cite{Roy_Steiner_04}, as well as many investigations \cite{Semileptonic_Flynn_07}-\cite{Lohse_90} that tackle different  aspects of this process and its connections to various issues in low-energy QCD.     Particularly,  the properties  of  scalar resonances in the $\pi K$ channel (the well established $K_0^*(1430)$ as well as the status of kappa meson)  have led to interesting studies in the field \cite{p1,p2,p3,p4,p6,p7,p8}.

Our main motivation for the study of $\pi K$ scattering in the present work is to extract information about scalar mesons in general,  and their underlying mixings,  in particular. Specifically,  we present a detailed analysis of the prediction of the generalized linear sigma model of Ref. \cite{global} for  $I$=1/2, 3/2 and  $J=0$ projection of the $\pi K$ scattering amplitude (and phase shifts).  Using K-matrix unitarization of the scattering amplitude we compare the predictions of the model with experimental data and analyze the poles of the scattering amplitude.   We also compute the scattering lengths in the $I$ = 1/2, 3/2 channels and compare with other studies.

After defining our  set up and notation in Sec. II, for orientation we start with a general discussion  of  the experimental data on $\pi K$ scattering amplitude in  Sec. III in which we explore the simplest mathematical structures for the bare amplitude  that, when unitarized with  the K-matrix method,  can fit the data.   To examine the effect of underlying mixing among scalar mesons,  we first give in Sec. IV the prediction of the single nonet SU(3) linear sigma model for the $\pi K$ scattering amplitude, and then after a short review of the generalized linear sigma model in Sec. V give its predictions for the scattering amplitudes and scattering lengths  in Sec. VI.   We conclude  with a  summary of the results in Sec. VII.

\section{Basic set up and notation}

\begin{figure}[!htb]
\centering
\includegraphics[scale=0.5]{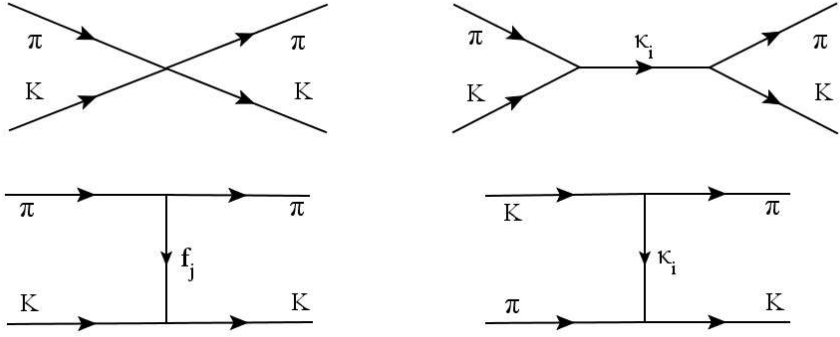}
\caption{Feynman diagrams representing the $\pi K$ scattering.    The parameters $n_f$ and $n_\kappa$ represent the number of isosinglet and isodoublet scalars respectively, which in the single (double) nonet model are equal to two (four) and one (two). }
\label{F_FD}
\end{figure}

The generic Feynman diagrams for this scattering are displayed in Fig. \ref{F_FD}.     In the single (double) nonet model
there are two (four) isosinglet scalars and one (two) isodoublet scalar(s) contributing to these diagrams.    For simplicity we ignore the vector meson contributions, but do not expect the results to change much based on comparison with prior work by some of the authors within a nonlinear chiral Lagrangian framework \cite{BFSS_98},  in which it was shown that while the effects of $\rho$ and $K^*$ are not individually negligible, their total effect is balanced by the vector contact contribution (see Fig. 3 of \cite{BFSS_98}).

The $I=3/2$ amplitude can be easily calculated from
\begin{equation}
A^{{3\over 2}}(s,t,u) = A \left( \pi^+(p_1) K^+(p_2) \rightarrow \pi^+(p_3) K^+(p_4) \right),
\end{equation}
which results in
\begin{eqnarray}
A^{\frac{3}{2}}(s,t,u)= -\gamma_{\pi K}^{(4)}+\sum_{i=1}^{n_{\kappa}} \frac{\gamma_{\kappa_i \pi K}^2}{m_{\kappa_i}^2-u}+\sum_{j=1}^{n_f} \frac{\gamma_{f_{j}KK}\gamma_{f_{j}\pi\pi}}{m_{f_j}^2-t},
\label{E_Astu}
\end{eqnarray}
where in the single (double) nonet model $n_f$ is two (four) and $n_\kappa$ is one (two), and the couplings are defined
by the Lagrangian density
\begin{equation}\label{relpot}
-{\cal L} = \gamma_{\pi K}^{(4)}{\bar K} K {\bm \pi} \cdot {\bm \pi} + {\gamma_{\kappa_i \pi K }\over \sqrt{2}}\left({\bar K} {\bm \tau} \cdot {\bm \pi} \kappa_i + {\rm H.c.} \right)
+ {\gamma_{f_j \pi \pi}\over \sqrt{2}} f_j {\bm \pi} \cdot {\bm \pi} + {\gamma_{f_j K K}\over \sqrt{2}}f_j {\bar K} K + \cdots.
\end{equation}
The $I=1/2$, $J=0$ scattering amplitude is needed
in order to investigate the properties of the $\kappa$ and $ K_0^*(1430)$ resonances in
the direct channels. The tree level amplitude involves $\kappa$ and $ K_0^*(1430)$  exchanges
in the $s$ and $u$ channels, $f_i$ exchanges in the
$t$ channel as well as a four point contact term. The relevant
tree level invariant amplitude may be written as
\begin{eqnarray}
A^{\frac{1}{2}}(s,t,u)=\frac{3}{2} A^{\frac{3}{2}}(u,t,s)-\frac{1}{2}A^{\frac{3}{2}}(s,t,u),
\end{eqnarray}
which results in
\begin{eqnarray}\label{a}
A^{\frac{1}{2}}(s,t,u)= -\gamma_{\pi K}^{(4)}+\frac{3}{2}\sum_{i=1}^{n_{\kappa}} \frac{\gamma_{\kappa_i  \pi K }^2}{m_{\kappa_i}^2-s}-\frac{1}{2}\sum_{i=1}^{n_{\kappa}} \frac{\gamma_{\kappa_i\pi K}^2}{m_{\kappa_i}^2-u}+\sum_{j=1}^{n_f} \frac{\gamma_{f_{j}KK}\gamma_{f_{j}\pi\pi}}{m_{f_j}^2-t}.
\end{eqnarray}
The $J=0$ partial wave amplitude is obtained from
\begin{eqnarray}
T_0^{\frac{1}{2}B}=\frac{\rho(s)}{2}\int_{-1}^1 d \cos\theta P_0(\cos \theta)A^\frac{1}{2}(s,t,u),
\label{T012B_def}
\end{eqnarray}
with $\rho(s)=q /(8\pi\sqrt{s})$ where $q$ is the center of mass momentum $q=\sqrt{(s-(m_{\pi}+m_K)^2)(s-(m_{\pi}-m_K)^2)}/(2\sqrt{s})$.
Performing the partial wave projection we find the ``bare'' $I=1/2,\, 3/2$, $J=0$ amplitudes
\begin{eqnarray}
T_0^{\frac{1}{2}B}&=&\frac{\rho (s)}{2}\left[-2\gamma_{\pi K}^{(4)}+3\sum_{i=1}^{n_{\kappa}} \frac{\gamma_{\kappa_i \pi K}^2}{m_{\kappa_i}^2-s}
-\frac{1}{4q^2}\sum_{i=1}^{n_{\kappa}} {\gamma_{\kappa_i\pi K}^2}\ln{\left(\frac{B_{\kappa_i}+1}{B_{\kappa_i}-1}\right)}
+\frac{1}{2q^2}\sum_{j=1}^{n_f} \gamma_{f_{j}KK}\gamma_{f_{j}\pi\pi}\ln{\left(1+\frac{4q^2}{m_{f_j}^2}\right)}\right],
\label{T012_B}\\
T_0^{\frac{3}{2}B}&=&\frac{\rho (s)}{2}\left[-2\gamma_{\pi K}^{(4)}+\frac{1}{2q^2}\sum_{i=1}^{n_{\kappa}} {\gamma_{\kappa_i\pi K}^2}\ln{\left(\frac{B_{\kappa_i}+1}{B_{\kappa_i}-1}\right)}+\frac{1}{2q^2}\sum_{j=1}^{n_f} \gamma_{f_{j}KK}\gamma_{f_{j}\pi\pi}\ln{\left(1+\frac{4q^2}{m_{f_j}^2}\right)}\right],
\label{T032_B}
\end{eqnarray}
in which
\begin{eqnarray}
B_{\kappa_i}=\frac{1}{2q^2}\left[(m_{\kappa_i})^2-m_K^2-m_{\pi}^2+2\sqrt{(m_{\pi}^2+q^2)(m_K^2+q^2)}\right],
\end{eqnarray}
and the Mandelstam variables are expressed in terms of $q$ and $\theta$
\begin{eqnarray}
t &=& -2 q^2 (1 - \cos \theta),
\nonumber \\
u &=&  m_{\pi}^2+m_K^2 - 2\sqrt{(m_{\pi}^2+q^2)(m_K^2+q^2)}-2q^2 \cos{\theta}.
\end{eqnarray}
Equations (\ref{T012_B}) and (\ref{T032_B}) are the ``Ansatz'' equations in this work.

Of course, the ``bare'' amplitude diverges at the ``bare'' isodoublet masses and should be regularized in some fashion, for example by adding imaginary parts in the denominator of the propagators,  or by other methods.     In the prior work on the prediction of this model for the $\pi\pi$ scattering amplitude and investigation of the unitarity corrections due to the final state interactions of pions,   the rather simple K-matrix method was used.  This method has the advantage of unitarizing the amplitude without introducing any new parameters, and therefore directly maps the Lagrangian parameters onto the scattering data.   This is particularly useful from the standpoint of global study of scalar mesons below and above 1 GeV where different mixing patterns in the Lagrangian are studied and any additional arbitrary parameters that may be introduced in the process of the unitarization can smear the probe of these important constants.    Since the  present study is conducted within the same global picture,  it is important for us to treat the unitarization procedure in the same way.   Hence, we employ the K-matrix unitarization in this work as well.   Having fixed all the parameters in the global picture of Ref. \cite{global},   and using K-matrix method,  we have no free parameters to fit for and therefore no flexibilities in comparing our prediction for the scattering amplitude with the experimental data.    The K-matrix unitarized amplitude is defined in terms of the ``bare'' amplitude by
\begin{eqnarray}\label{unt}
T_0^{\frac{1}{2}}=\frac{T_0^{\frac{1}{2}B}}{1-i T_0^{\frac{1}{2}B}}
\label{T012_unitary}
\end{eqnarray}
This is what we take as our physical amplitude and  compare with the experimental data.

The physical masses (${\widetilde m}_i$) and decay widths (${\widetilde \Gamma}_i$) are determined from the poles in the unitarized amplitude.   Solving for the roots ($z_i$) of the denominator of (\ref{T012_unitary})
\begin{equation}
1-i T_0^{\frac{1}{2}B} = 0 \Rightarrow z_i = {\widetilde m}_i^2 - i {\widetilde m}_i {\widetilde \Gamma}_i.
\label{poles_eq}
\end{equation}
In general,  some of the poles may not be physical (for example, being below the threshold).

\section{General treatment of $\pi K$ scattering data using $K$-matrix}

\begin{figure}[!htb]
\begin{center}
\vskip 1cm
\epsfxsize = 7.5cm
 \includegraphics[width=8cm]{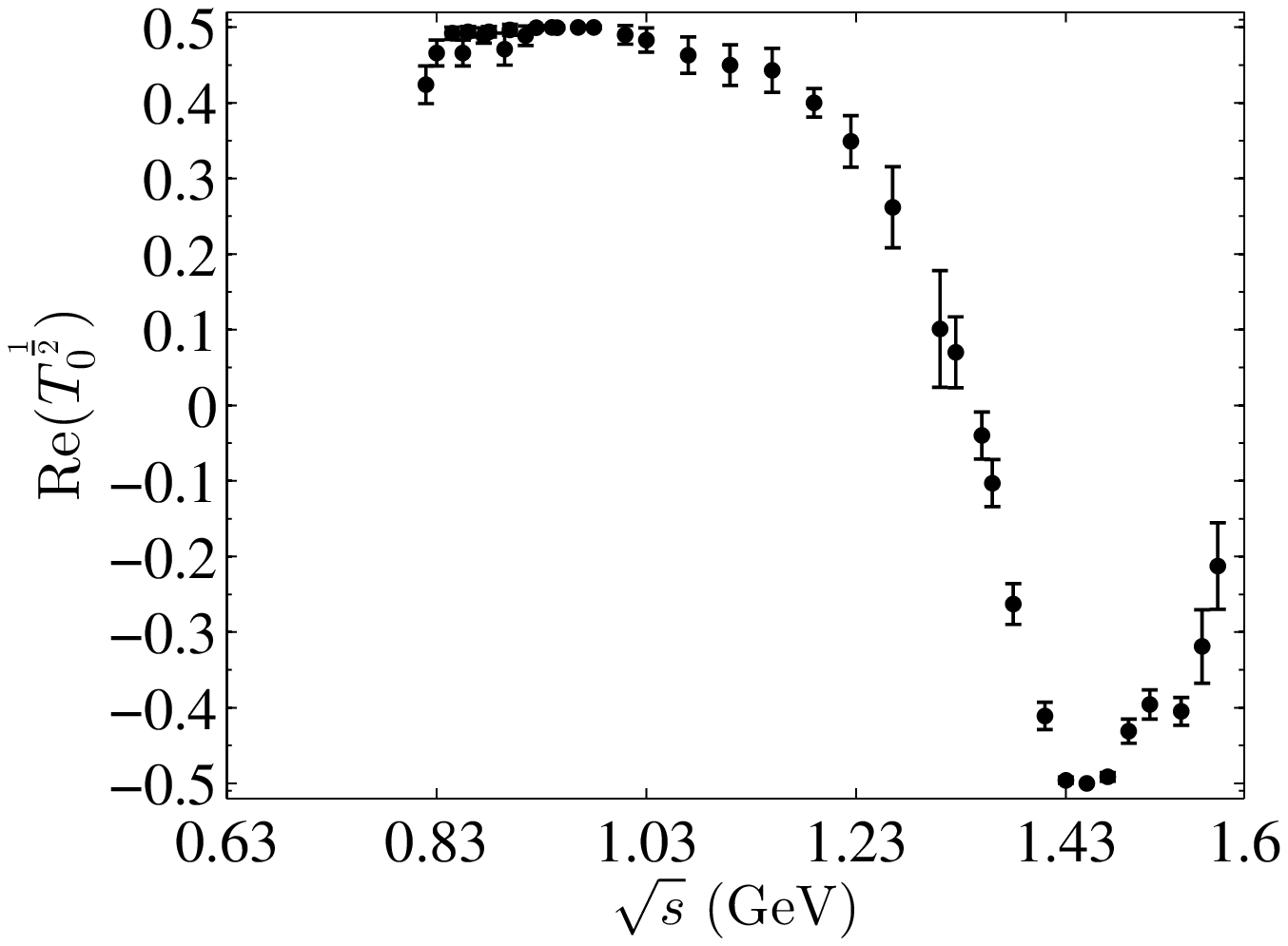}
\hskip 1cm
\epsfxsize = 7.5cm
 \includegraphics[width=8cm]{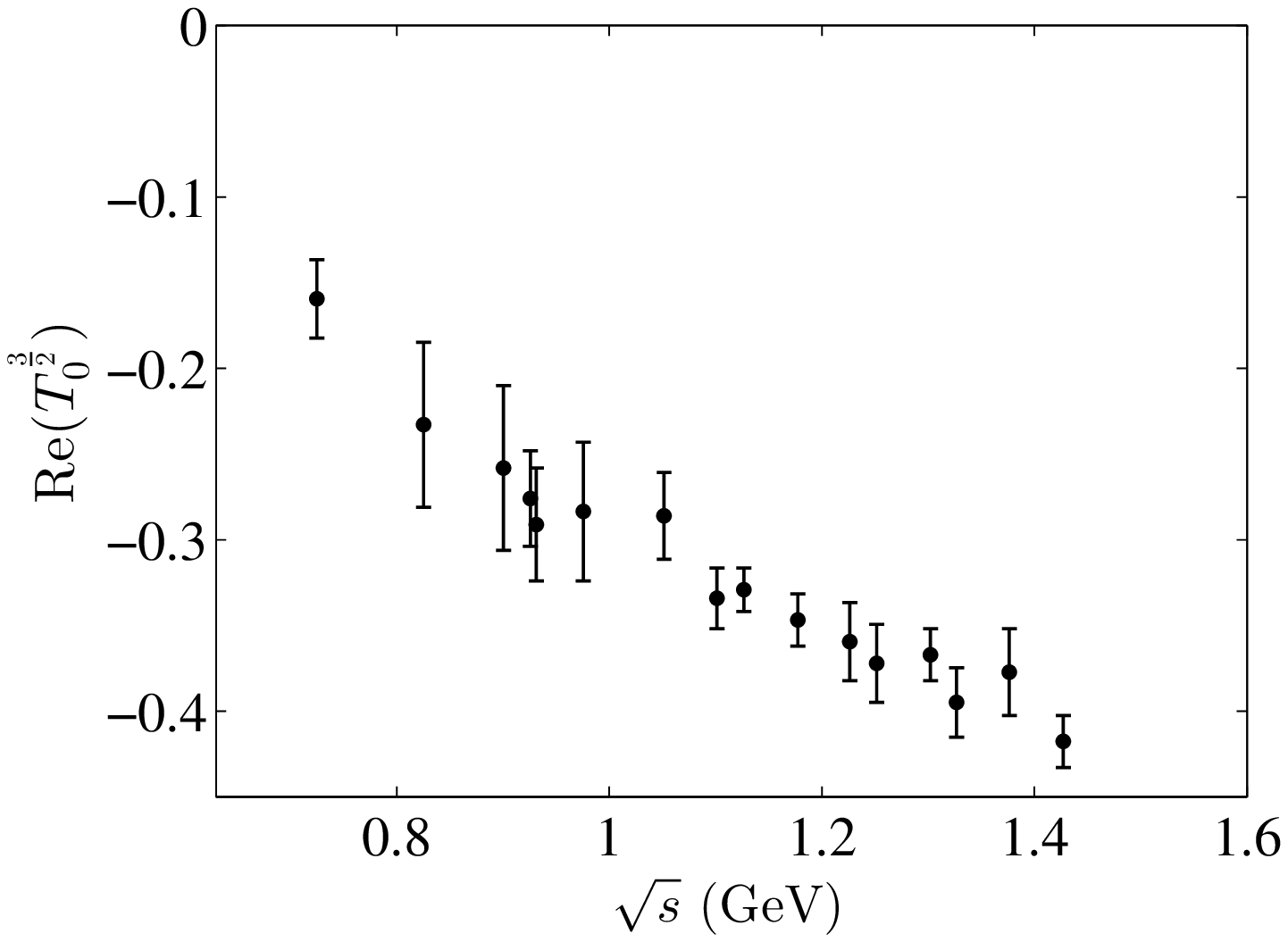}
 \caption{The real part of the $J=0$, $I=1/2$ (left) and $I=3/2$ (right), $\pi K$ scattering amplitudes extracted from the experimental data of Ref. \cite{Aston}. }
\label{data}
\end{center}
\end{figure}

For orientation,  in this section  we study the $\pi K$ scattering data using K-matrix method and explore the mathematical structures for the bare amplitude that can fit the data and the resulting physical parameters that can be consequently inferred.    The experimental data on $\pi K$ scattering amplitude in $I=1/2$, $J=0$ channel  and in $I=3/2$, $J=0$ channel are extracted from the phase shift data of Ref.\cite{Aston} and given in Fig. \ref{data}.
For the case of $I=1/2$, $J=0$,  we see that  the data vanishes around 1.3 GeV followed by a dip around 1.45 GeV and beyond that the data steadily increases (due to lack of data beyond 1.6 GeV it is not clear whether the real part of the amplitude approaches zero in this region but this seems to be a plausible possibility).   Therefore, if the K-matrix unitarization is to provide a reasonable description of data,    the
physical amplitude must vanishes around 1.3 GeV (and possibly somewhere above 1.6 GeV).
We note that the K-matrix unitarized amplitude vanishes at points where the bare amplitude either vanishes or has a pole, i.e.
\begin{eqnarray}
 T_0^{{1\over 2}B} \rightarrow 0 \hskip 0.3cm &\Rightarrow& \hskip 0.3cm T_0^{{1\over 2}} \rightarrow 0, \nonumber \\
 T_0^{{1\over 2}B} \rightarrow \infty \hskip 0.3cm &\Rightarrow& \hskip 0.3cm T_0^{{1\over 2}}\rightarrow  0.
\label{bare_amp_cond}
\end{eqnarray}
This means that in order to describe the data using K-matrix,  the bare amplitude should either diverge or vanish around 1.3 GeV (and possibly around 1.6 GeV).   These general guidelines help examining  the possible structures for the bare amplitude.

\begin{figure}[!htb]
	\begin{center}
		\epsfxsize = 7.5cm
		\includegraphics[width=8cm]{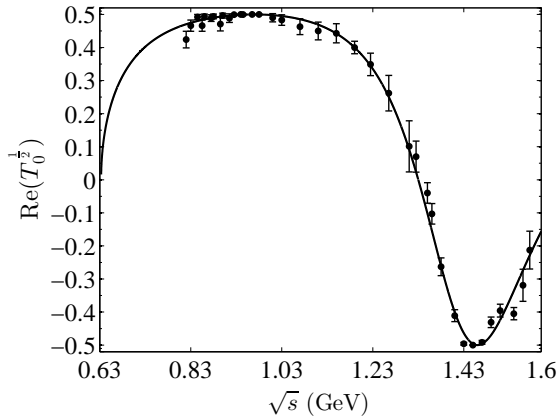}
		\caption{Fit of the K-matrix unitarized simple toy structure (\ref{simplepole}) to experimenal data.  }
		\label{F_mod-Ind_1}
	\end{center}
\end{figure}

The simplest mathematical structure for the bare amplitude that fits $I=1/2$, $J=0$ data consists of a constant background and a pole
\begin{equation}
T_0^{{1\over 2}B} =\frac{\rho(s)}{2}\left[c_0 + \frac{c}{m^2-s}\right]
\label{simplepole}
\end{equation}
Substituting the ``bare'' amplitude  into Eq. (\ref{T012_unitary}) the physical amplitude is obtained.   We then fit the real part of the physical amplitude to the data displayed in Fig. \ref{data} and find that for the free parameters $c_0\approx 64$, $c\approx 65\, {\rm GeV}^2$ and $m\approx 1.33 \, {\rm GeV}$) we get a good description of data up to about 1.6 GeV (Fig. \ref{F_mod-Ind_1}).   In this case,  the bare amplitude vanishes around 1.3 GeV which agrees with the intuitive understanding of data using property (\ref{bare_amp_cond}).    The pole in the bare amplitude yields the first vanishing point in the unitarized amplitude.   In addition,  the interference of the background and the pole results in  vanishing of the  bare amplitude around   1.6 GeV, which in turn leads to a vanishing point in the unitarizd amplitude around 1.6 GeV.

Moreover,  we examine the pole(s) of the physical amplitude  by solving for the complex roots of the  denominator of the K-matrix unitarized amplidue   (\ref{poles_eq}) and find the physical mass and decay width
\begin{eqnarray}
{\widetilde{m}} &=& 1447 \,\, {\rm MeV}, \nonumber \\
{\widetilde {\Gamma}} &=& 313 \,\, {\rm MeV},
\end{eqnarray}
which are clearly the mass and total decay width of $K_0^*(1430)$ with experimental values  \cite{pdg}:
\begin{eqnarray}
m[K_0^*(1430)]&=&1424\pm 50\,\, {\rm MeV} \nonumber \\
\Gamma[K_0^*(1430)]&=& 270\pm 80 \,\, {\rm  MeV}
\end{eqnarray}

\begin{figure}[!htb]
	\begin{center}
		\epsfxsize = 7.5cm
		\includegraphics[width=8cm]{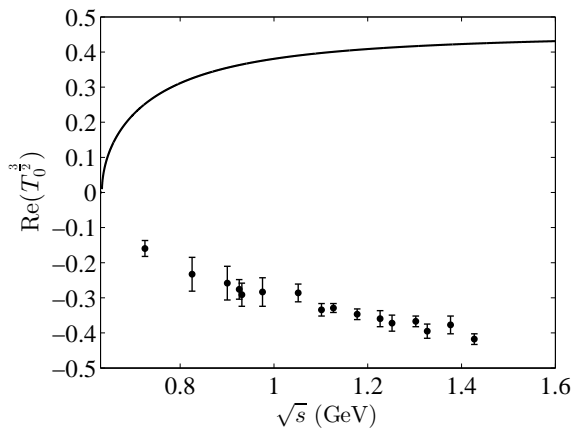}
		\caption{The $I=3/2$, $J=0$, $\pi K$  scattering amplitude (solid line) obtained with the same fitted parameters  of Fig. \ref{F_mod-Ind_1} is compared with experimental data.}
		\label{F_mod-Ind_2}
	\end{center}
\end{figure}

The simple structure (\ref{simplepole})  accomplishes two important objectives:  (a) It fits the experimental data on the real part of $I=1/2$, $J=0$ up to about 1.6 GeV;  and (b) The poles of the physical amplitude give the mass and the decay width of $K_0^*(1430)$ showing that the $I=1/2$ and $J=0$  channel is dominated by the effect of this resonance.     However, there are two objectives that have not been met: (i) When the same fitted parameters are used to compute the  $I=3/2$, $J=0$, the result is far from experimental data (Fig. \ref{F_mod-Ind_2}); and (ii) The kappa meson is left undetected.       To gain additional insight from the experimental data we have examined more involved mathematical structures.    These include adding $u$- and $t$-channel type contributions [see Eqs. (\ref{E_Astu}), (\ref{T012_B})], as well as the second pole (perhaps coming from a second scalar nonet).    We find that all such structures again fit the  real part of $I=1/2$, $J=0$ well, but still are not able to simultaneously fit  the $I=3/2$, $J=0$ amplitude, leaving issue (i) above unresolved.     Regarding detection of kappa meson [issue (ii) above] we find that the interplay of background and resonance(s) play a very important role in detecting kappa.     Among all the  test structures that we have used to fit the real part of the  $I=1/2$, $J=0$  amplitude,  only those that include the $t$-channel contributions yield two physical poles, one of which is always the $K_0^*(1430)$ and the second  one results in a mass around 665 MeV and a decay width around 350 MeV which is close to the kappa meson's property.

In summary,  we saw in this Sec. that fitting to experimental data in a given channel (such as the $I=1/2$, $J=0$  channel) is easily achievable with a mathematical structure as simple as a constant background and a pole.   We also saw that although such structures are not unique, they all detect the $K_0^*(1430)$ in a very close agreement with experiment.   The nontrivial aspects which are not easily achievable with arbitrary structures are: simultaneously describing both the $I=1/2$ and $I=3/2$ channels as well as detecting the kappa meson.    We will see that the generalized linear sigma model employed in the present work, when applied below 1 GeV,  is able to  achieve both objectives (i) and (ii).



\section{$\pi K$ scattering in single nonet linear sigma model}


In order to study the effects of underlying two- and four-quark mixings of scalar mesons in $\pi K$ scattering we compare the predictions of the generalized linear sigma model (double nonet) with that obtained within single nonet linear sigma model in \cite{LsM}.    In this section we briefly present the single nonet results.    The three flavor linear sigma model is constructed from the $ 3\times3$ chiral field
\begin{equation}
M = S +i\phi,
\label{sandphi}
\end{equation}
where $S=S^{\dagger}$ represents a scalar nonet and $ \phi=\phi^{\dagger}$ a pseudoscalar nonet. Under a chiral transformation $q_L\rightarrow U_L q_L$,  $q_R\rightarrow U_R q_R$ of the fundamental left and right handed light quark fields, $M$ transforms as
\begin{equation}
M \rightarrow U_L M U_R^{\dagger}.
\label{sm}
\end{equation}
The Lagrangian density has the general structure
\begin{equation}
{\cal L} = - \frac{1}{2} {\rm Tr}
\left( \partial_\mu M \partial_\mu M^{\dagger}
\right) - V_0 \left( M \right) - V_{SB},
\label{LsMLag}
\end{equation}
where $V_0$ is an arbitrary function of the independent
$\rm{SU(3)_L} \times {\rm SU(3)_R} \times {\rm U(1)_V}$ invariants
\begin{eqnarray}
I_1&=&{\rm Tr} (M M^{\dagger}), \hspace{1cm} I_2={\rm Tr}(MM^{\dagger}MM^{\dagger}),\nonumber \\
I_3&=&{\rm Tr}\big((M M^{\dagger})^3\big)…,\hspace{.6 cm} I_4=6(\det M + \det M^{\dagger} ),
\end{eqnarray}
of which, only $I_4$ is not invariant under ${\rm U(1)_A}$. The symmetry
breaker $V_{SB}$ has the minimal form
\begin{equation}
V_{SB}=-2(A_1 S_1^1 + A_2 S_2^2 + A_3 S_3^3),
\end{equation}
with vacuum values satisfy
\begin{equation}
\langle S_a^b\rangle= \alpha_a \delta_a^b.
\end{equation}
The one-point vertices (pseudoscalar decay constants)
are related to these parameters by
\begin{equation}
F_{\pi}=\alpha_1+\alpha_2 ,  \hspace{1cm}  F_K=\alpha_1+\alpha_3,
\end{equation}
where in the isotopic spin invariant limit
\begin{equation}
A_1=A_2 ,  \hspace{1cm} \alpha_1=\alpha_2.
\end{equation}
We also need the minimum condition
\begin{equation}
\left< \frac{\partial V}{\partial S_a^b}\right>=0.
\end{equation}
The formula for the mass of the $\eta^{\prime}$ also involves the quantity
\begin{equation}
V_4\equiv\left< \frac{\partial V_0}{\partial I_4}\right>.
\end{equation}
Many of the three point and four point vertices may be
obtained by respectively two times and three times differentiating
the above mentioned generating equations \cite{LsM}. The five parameters, $A_1$ , $A_3$ , $\alpha_1$ , $\alpha_3$
and $V_4$ are determined by using the five experimental
inputs:
\begin{eqnarray}
m_{\pi} &=& 137\hspace{.1cm} {\rm MeV},\hspace{.2cm} m_K = 495 \hspace{.1cm}{\rm MeV},\nonumber \\
m_{\eta} &=& 547 \hspace{.1cm} {\rm MeV},\hspace{.2cm} m_{\eta^{\prime}} = 958 \hspace{.1cm} {\rm MeV},\nonumber \\
F_\pi &=& 131 \hspace{.1cm} {\rm MeV}
\label{inputs}
\end{eqnarray}
Masses of pseudoscalars are completely determined based on the underlying chiral symmetry together with the choice of symmetry breakers (both U(1)$_{\rm A}$ and SU(3)$_{\rm L} \times$ SU(3)$_{\rm R}$ $\rightarrow$ SU(2) isospin).    The scalar masses on the other hand are not all predicted;  in the most general case only the mass of isodoublet kappa meson is predicted, whereas
if the renormalizability is imposed the isovector mass and one of the isosinglet masses are determined.    It is found in \cite{LsM} that it is necessary not to  impose the renomalizability condition in order to be able to fit to the $\pi \pi$ scattering amplitude.   In the nonrenormalizable case, the ``bare'' scalar masses $m_{BARE}(\sigma)$, $m_{BARE}(f_0)$ and $m_{BARE}(a_0)$ (i.e. the Lagrangian masses which are different from the physical masses derived from the poles of the appropriate unitarized scattering amplitudes) and the scalar mixing angle $\theta_s$  are found from fits to various low-energy data in \cite{LsM}.    With the same set of parameters, the $I$=1/2, $J$=0, $\pi K$ scattering amplitude is obtained using our ``Ansatz'' equation (\ref{T012_B}).   The required coupling constants  are computed from the ``generating equations'' that express the symmetry of the Lagrangian  (\ref{LsMLag}) (an algorithm that facilitates such rather tedious computations is presented in \cite{LsM_Maple}):
\begin{eqnarray}
\gamma^{(4)}_{\pi K} &=& \left\langle \frac{\partial^4 V}{\partial \phi_1^2 \partial\phi_2^1\partial\phi_1^3\partial\phi_3^1}\right\rangle_0,\nonumber\\
\gamma_{\kappa \pi K}&=&\sum_a \left\langle \frac{\partial^3 V}{\partial\phi_1^2 \partial S_2^3 \partial\phi_3^1 }\right\rangle_0,  \nonumber\\
\gamma_{f_j \pi\pi}&=&{1\over \sqrt{2}}\,\sum_a \left\langle \frac{\partial^3 V}{\partial S_a^a \partial\phi_1^2 \partial\phi_2^1}\right\rangle_0 (R_s)_{j+1}^a, \nonumber\\
\gamma_{f_j K K}&=& \sqrt{2} \,\sum_a \left\langle \frac{\partial^3 V}{\partial S_a^a \partial\phi_1^3 \partial\phi_3^1}\right\rangle_0 (R_s)_{j+1}^a,
\end{eqnarray}
where the ``bare'' couplings and the rotation matrices ($R_s$ and $R_\phi$) are given in Appendix A.    Here $f_1=\sigma$ and $f_2=f_0(980)$.

\begin{figure}[!htb]
\begin{center}
\vskip 1cm
\epsfxsize = 7.5cm
 \includegraphics[width=8cm]{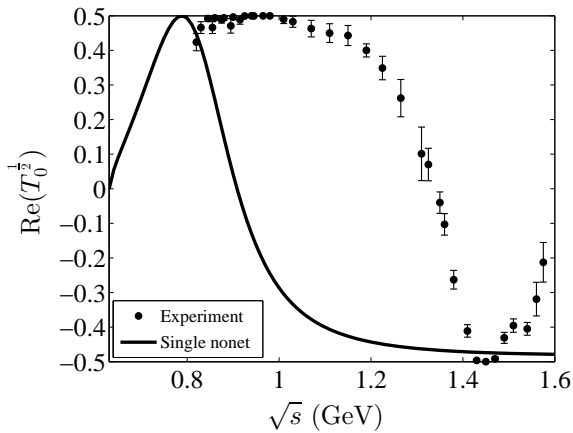}
 \caption{Real part of the $I=1/2$, $J=0$, $\pi K$ scattering amplitude predicted by the single nonet linear sigma model of Ref. \cite{LsM} (solid line) is compared with experimental data extracted from \cite{Aston}.}
\label{F_T012_sn}
\end{center}
\end{figure}

Using the inputs (\ref{inputs}) together with the result of best fit to pi pi scattering amplitude of Ref. \cite{LsM},
the bare $I$=1/2, $J=0$, $\pi K$ scattering amplitude is computed from Eq. (\ref{T012_B}) and K-matrix unitarized according to (\ref{T012_unitary}).   The result is plotted in Fig. \ref{F_T012_sn} and compared with experimental data extracted from \cite{Aston}.   Despite the success of SU(3) single nonet linear sigma model in describing $\pi\pi$ scattering up to about 1.2 GeV, we see that simultaneous description of $\pi K$ amplitude is not good.     This can be  attributed to the absence  of the second strange scalar $K_0^*(1430)$ in the single nonet model, where as we saw in Sec. III,  this resonance seems to dominate the $\pi K$ scattering data in $I=1/2$, $J=0$ channel. 
In the single nonet model of Ref. \cite{LsM},  with parameters that give a good description of $I=J=0$, $\pi \pi$ scattering data up to 1.2 GeV, there is only a pole in the $I=1/2$, $J=0$, $\pi K$ scattering amplitude at 0.847 GeV, and therefore understandable why the $\pi K$ description is not good within the single nonet approach (which is different than the $I=J=0$, $\pi\pi$  amplitude where the two isosinglet states of the single nonet model match well with the two points where the experimental data for the amplitude vanishes;  see Fig. 8 of \cite{LsM}).  In the generalized linear sigma model (double nonet model) there are two $I=1/2$ strange mesons,  and that, at least in principle,  may  make the description of $\pi K$ amplitude more feasible.

\section{Brief Review of the generalized linear sigma model}


In this section we give a brief review of the generalized linear sigma model of \cite{global} and references therein.  The model is constructed in terms of 3$\times$3 matrix
chiral nonet fields:
\begin{equation}
M = S +i\phi, \hskip 2cm
M^\prime = S^\prime +i\phi^\prime,
\label{sandphi}
\end{equation}
where $M$ and $M'$ transform in the same way under
chiral SU(3) transformations but transform differently under U(1)$_A$
transformation properties. $M$ describes the ``bare"
quark antiquark scalar and pseudoscalar nonet fields while
$M'$ describes ``bare" scalar and pseudoscalar fields
containing two quarks and two antiquarks.
The exact substructure of $M'$ is not probed by the model and in general
can be a linear combination of a diquark-antidiquark and a molecular structure.
The model distinguishes  $M$  from $M'$ through the U(1)$_A$ transformation.

The Lagrangian density has the general structure
\begin{equation}
{\cal L} = - \frac{1}{2} {\rm Tr}
\left( \partial_\mu M \partial_\mu M^\dagger
\right) - \frac{1}{2} {\rm Tr}
\left( \partial_\mu M^\prime \partial_\mu M^{\prime \dagger} \right)
- V_0 \left( M, M^\prime \right) - V_{SB},
\label{mixingLsMLag}
\end{equation}
where $V_0(M,M^\prime)$ stands for a function made
from SU(3)$_{\rm L} \times$ SU(3)$_{\rm R}$
(but not necessarily U(1)$_{\rm A}$) invariants
formed out of $M$ and $M^\prime$.  In principle,  there are infinite number of such terms.   Even if
we only consider the renormalizable terms,  there are still 21 terms that can be written down for
$V_0$.     To keep the calculations in this model tractable, it is practical to define an approximation scheme
that allows limiting the number of terms at each level of calculation, and systematically improving the results thereafter.
Such a scheme was defined in \cite{07_FJS1}, in terms of the number of the quark and antiquark in each term.
The  leading choice of terms
corresponding
to eight or fewer underlying quark plus antiquark lines
 at each effective vertex
reads:
\begin{eqnarray}
V_0 =&-&c_2 \, {\rm Tr} (MM^{\dagger}) +
c_4^a \, {\rm Tr} (MM^{\dagger}MM^{\dagger})
\nonumber \\
&+& d_2 \,
{\rm Tr} (M^{\prime}M^{\prime\dagger})
     + e_3^a(\epsilon_{abc}\epsilon^{def}M^a_dM^b_eM'^c_f + {\rm H. c.})
\nonumber \\
     &+&  c_3\left[ \gamma_1 {\rm ln} (\frac{{\rm det} M}{{\rm det}
M^{\dagger}})
+(1-\gamma_1){\rm ln}\frac{{\rm Tr}(MM'^\dagger)}{{\rm
Tr}(M'M^\dagger)}\right]^2.
\label{SpecLag}
\end{eqnarray}
     All the terms except the last two (which mock up the axial anomaly)
      have been chosen to also
possess the  U(1)$_{\rm A}$
invariance.   A possible term $\left[{\rm Tr} (M M^{\dagger})\right]^2$ is neglected for simplicity because it violates the OZI rule.
The symmetry breaking term which models the QCD mass term
takes the form:
\begin{equation}
V_{SB} = - 2\, {\rm Tr} (A\, S),
\label{vsb}
\end{equation}
where $A=$diag$(A_1,A_2,A_3)$ are proportional to
  the three light quark
masses.
The model allows for two-quark condensates,
$\alpha_a=\langle S_a^a \rangle$ as well as
four-quark condensates
$\beta_a=\langle {S'}_a^a \rangle$.
Here we assume \cite{SU} isotopic spin
symmetry so A$_1$ =A$_2$ and:
\begin{equation}
\alpha_1 = \alpha_2  \ne \alpha_3, \hskip 2cm
\beta_1 = \beta_2  \ne \beta_3.
\label{ispinvac}
\end{equation}
 We also need the ``minimum" conditions,
\begin{equation}
\left< \frac{\partial V_0}{\partial S}\right> + \left< \frac{\partial
V_{SB}}{\partial
S}\right>=0,
\quad \quad \left< \frac{\partial V_0}{\partial S'}\right>
=0.
\label{mincond}
\end{equation}
There are twelve parameters describing the Lagrangian and the
vacuum. These include the six coupling constants
 given in Eq.(\ref{SpecLag}), the two quark mass parameters,
($A_1=A_2,A_3$) and the four vacuum parameters ($\alpha_1
=\alpha_2,\alpha_3,\beta_1=\beta_2,\beta_3$). The four minimum
equations reduce the number of needed input parameters to
eight.

Five of these eight are supplied by the following
masses together with the pion decay constant:
\begin{eqnarray}
 m[a_0(980)] &=& 980 \pm 20\, {\rm MeV},
\nonumber
\\ m[a_0(1450)] &=& 1474 \pm 19\, {\rm MeV},
\nonumber \\
 m[\pi(1300)] &=& 1300 \pm 100\, {\rm MeV},
\nonumber \\
 m_\pi &=& 137 \, {\rm MeV},
\nonumber \\
F_\pi &=& 131 \, {\rm MeV}.
\label{inputs1}
\end{eqnarray}
Because $m[\pi(1300)]$ has such a large uncertainty,
we will, as previously, examine predictions
depending on the choice of this mass
within its experimental range.
The sixth input will be taken as the light
``quark mass ratio" $A_3/A_1$, which will
be varied over an appropriate range.
 The remaining two inputs (to be traded with $c_3$ and $\gamma_1$) are taken from the
 masses of the  pseudoscalar
mesons \cite{global}.    However,  $c_3$ and $\gamma_1$ affect the $\eta$ system only and are not needed in the present study.
Given these inputs a  number of
predictions are made in \cite{global}. At the level of the quadratic terms in the
Lagrangian, we predict all the remaining masses
 and decay constants as well
as the angles describing the mixing between each of
($\pi,\pi'$),
($K,K'$), ($a_0,a_0'$), ($\kappa,\kappa'$) multiplets
and each of the 4$\times$4
isosinglet mixing matrices
 (each formally described by six angles).

Consequently,  all twelve parameters of the model (at the present order of approximation) are evaluated by the method discussed above using four minimum equations and eight experimental inputs.   The uncertainties of the experimental inputs result in uncertainties on the twelve model parameters which in turn result in uncertainties on  physical quantities that are computed in this model.    In the work of Ref. \cite{global} all rotation matrices describing the underlying mixing among two- and four-quark components for each spin and isospin states are computed.
For the study of $\pi K$ scattering, we need the following rotation matrices:
\begin{equation}
\left[
\begin{array}{cc}
K_0(800)\\
K_0^*(1430)
\end{array}
\right]
=
L_\kappa^{-1}
\left[
\begin{array}{cc}
S_1^3\\
{S'}_1^3
\end{array}
\right],
\hskip 2cm
\left[
\begin{array}{cc}
f_1\\
f_2\\
f_3\\
f_4
\end{array}
\right]
=
L_0^{-1}
\left[
\begin{array}{cc}
f_a\\
f_b\\
f_c\\
f_d
\end{array}
\right],
\label{E_SRot}
\end{equation}
where $L_\kappa^{-1}$ and  $L_a^{-1}$ are the rotation matrices for  $I=1/2$ and $I=0$ scalars respectively; $f_i, i=1..4$ are four of the physical isosinglet scalars below 2 GeV (in this model $f_1$ and $f_2$ are clearly identified with $f_0(500)$ and $f_0(980)$ and the two heavier states resemble two of the heavier isosinglet scalars above 1 GeV); and
\begin{eqnarray}
f_a&=&\frac{S^1_1+S^2_2}{\sqrt{2}} \hskip .7cm
\propto n{\bar n},
\nonumber  \\
f_b&=&S^3_3 \hskip 1.6 cm \propto s{\bar s},
\nonumber    \\
f_c&=&  \frac{S'^1_1+S'^2_2}{\sqrt{2}}
\hskip .5 cm \propto ns{\bar n}{\bar s},
\nonumber   \\
f_d&=& S'^3_3
\hskip 1.5 cm \propto nn{\bar n}{\bar n},
\label{f_basis}
\end{eqnarray}
where the non-strange ($n$) and strange ($s$) quark content
for each basis state has been listed at the end of
each line above.
For pseudoscalars:
\begin{equation}
\left[
\begin{array}{cc}
\pi^+(137)\\
\pi^+(1300)
\end{array}
\right]
=
R_\pi^{-1}
\left[
\begin{array}{cc}
\phi_1^2\\
{\phi'}_1^2
\end{array}
\right],
\hskip 2cm
\left[
\begin{array}{cc}
K^+(496)\\
{K'}^+(1460)
\end{array}
\right]
=
R_K^{-1}
\left[
\begin{array}{cc}
\phi_1^3\\
{\phi'}_1^3
\end{array}
\right],
\label{E_PRot}
\end{equation}
where $R_\pi^{-1}$ and $R_K^{-1}$ are the rotation matrices for $I=1$ and $I=1/2$ pseudoscalars respectively.


\section{Generalized linear sigma model prediction of $\pi K$ scattering Amplitude}

 The Feynman diagrams for this scattering are given in Fig. \ref{F_FD}  and include a four-point contact term, contribution of isodoublet scalars in the $s$- and $u$-channels, as well as contribution of isosinglet scalars in the $t$-channel.   It is shown in \cite{BFSS1} that the total contribution of vectors are negligible compared to other contributions.
 The part of potential relevant to this investigation is given in Eq. (\ref{relpot})
in which the coupling constants are
\begin{eqnarray}
\gamma_{\pi K}^{(4)}&=&\left\langle \frac{\partial^4 V}{\partial \pi^{+} \partial \pi^{-} \partial K^{+}  \partial K^{-}} \right\rangle
= \sum_{A,B,C,D} \left\langle \frac{\partial^4 V}{\partial (\phi_1^2) \partial (\phi_2^1) \partial (\phi_1^3)  \partial (\phi_3^1)} \right\rangle (R_\pi)_{A1} \, (R_\pi)_{B1} \, (R_K)_{C1} \, (R_K)_{D1},
\nonumber \\
\gamma_{f_j\pi\pi} &=&
{1\over {\sqrt{2}}}
\left\langle
{{\partial^3 V}
\over
{\partial f_j \, \partial \pi^+ \, \partial \pi^-}}
\right\rangle
=
{1\over {\sqrt{2}}}
\sum_{I,A,B}
\left\langle
{{\partial^3 V}
\over
{
 \partial f_I \,
 \partial (\phi_1^2)_A \,
 \partial (\phi_2^1)_B
}}
\right\rangle
(L_0)_{I i} \,
(R_\pi)_{A1} \,
(R_\pi)_{B1},
\nonumber \\
\gamma_{f_j K K} &=&
\sqrt{2}\left\langle \frac{\partial^3V}{\partial f_j\partial K^{+}\partial K^{-}} \right\rangle=\sqrt{2}
\sum_{I,A,B}\frac{\partial^3V}{\partial f_I\partial(\phi_1^3)_A\partial (\phi_3^1)_B}(L_0)_{Ii}(R_K)_{A1}(R_K)_{B1},
\nonumber \\
\gamma_{\kappa_i \pi K} &=&
\left\langle
{{\partial^3 V}
\over
{\partial \kappa_i^0 \, \partial K^- \, \partial \pi^+}}
\right\rangle
=
\sum_{A,B,C}
\left\langle
{{\partial^3 V}
\over
{
 \partial (S_2^3)_A \,
 \partial  (\phi_3^1)_B \,
 \partial (\phi_1^2)_C
 }}
\right\rangle
(L_\kappa)_{Ai} \,
(R_K)_{B1} \,
(R_\pi)_{C1},
\end{eqnarray}
where $A$, $B$  and $C$ can take values of 1 and 2 (with 1 referring to nonet $M$ and 2 referring to nonet $M'$) and $I$ is a placeholder for  {\it a},{\it b},{\it c} and {\it d} that
represent the four bases in Eq.
(\ref{f_basis}).  $L_0$, $L_\kappa$, $R_0$, $R_K$, $R_\pi$  are the rotation matrices defined in previous Sec. V.  The bare coupling constants  are all given in Appendix B.    We also show in Appendix C how the present framework recovers the current algebra result.

\subsection{Comparison of $I=1/2$ channel with experiment}
\begin{figure}[!htb]
\begin{center}
\vskip 1cm
\epsfxsize = 7.5cm
 \includegraphics[width=8cm]{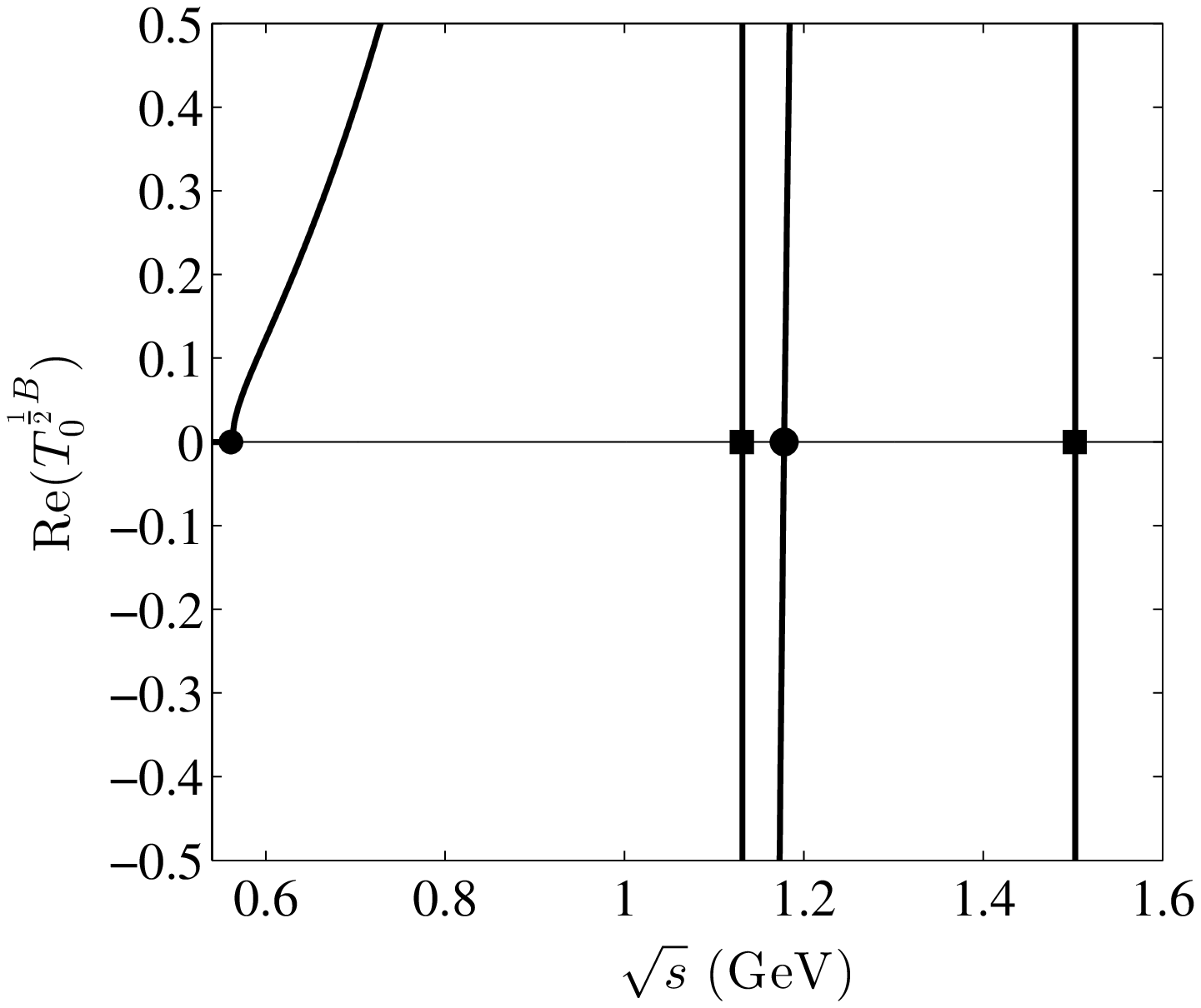}
\hskip 1cm
\epsfxsize = 7.5cm
 \includegraphics[width=8cm]{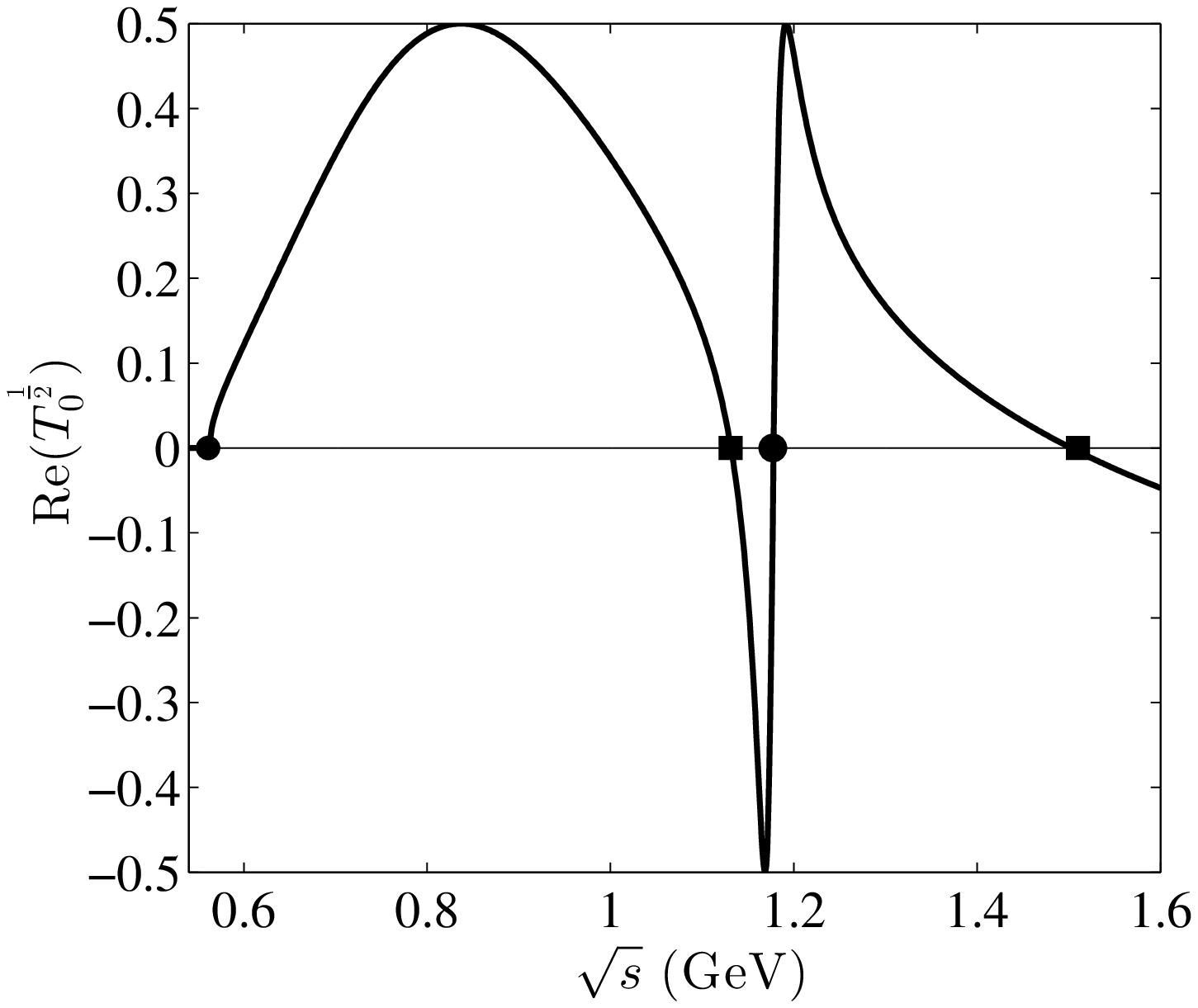}
 \caption{Real part of the $I=1/2$, $J=0$, $\pi K$ scattering amplitude.     The bare amplitude (left) contains zeros (circles) and poles (squares) at which the unitarized amplitude (right) vanishes.}
\label{F_generic_T120}
\end{center}
\end{figure}

\begin{figure}[!htb]
\begin{center}
\vskip 1cm
\epsfxsize = 7.5cm
 \includegraphics[width=8cm]{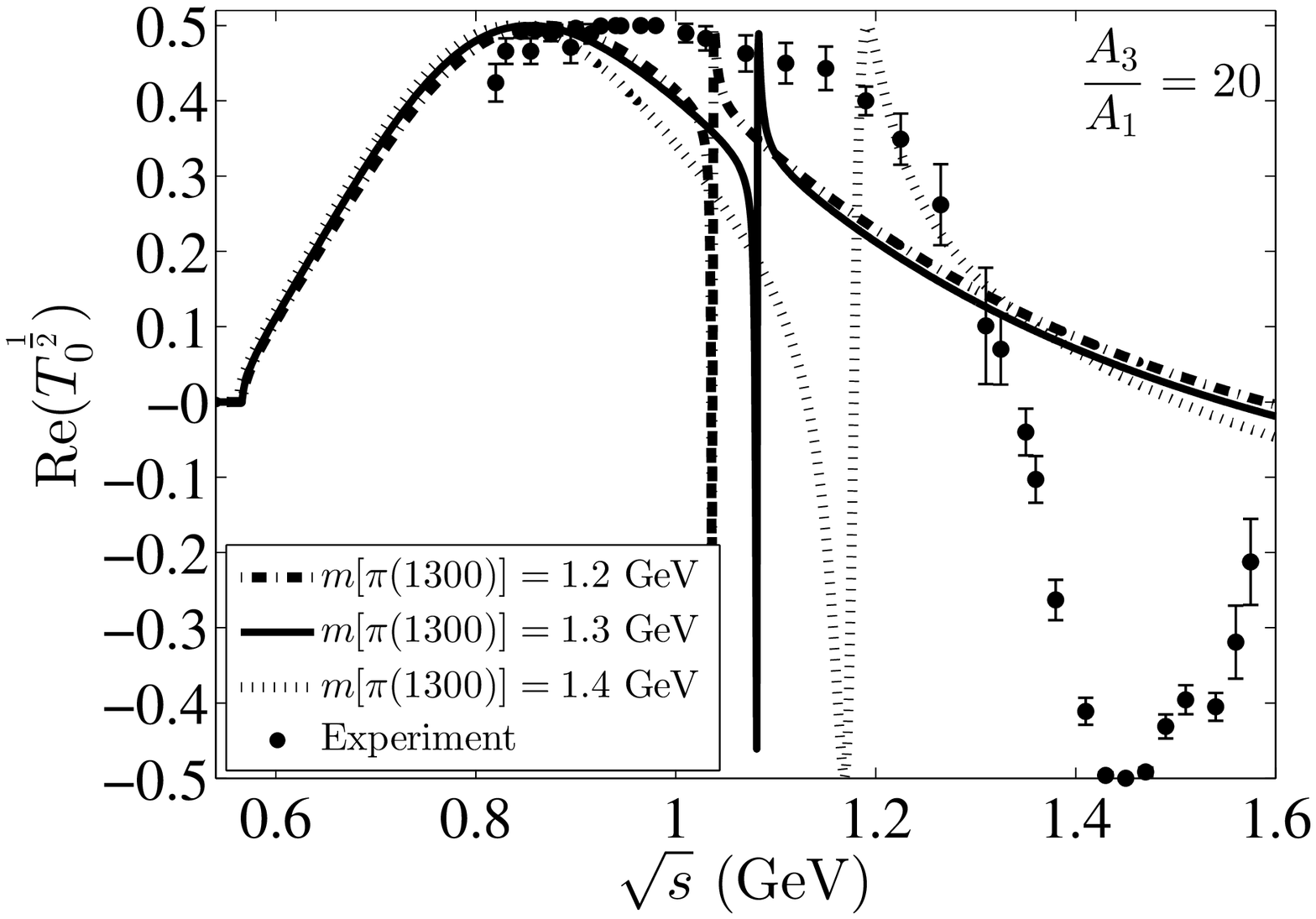}
\hskip 1cm
\epsfxsize = 7.5cm
 \includegraphics[width=8cm]{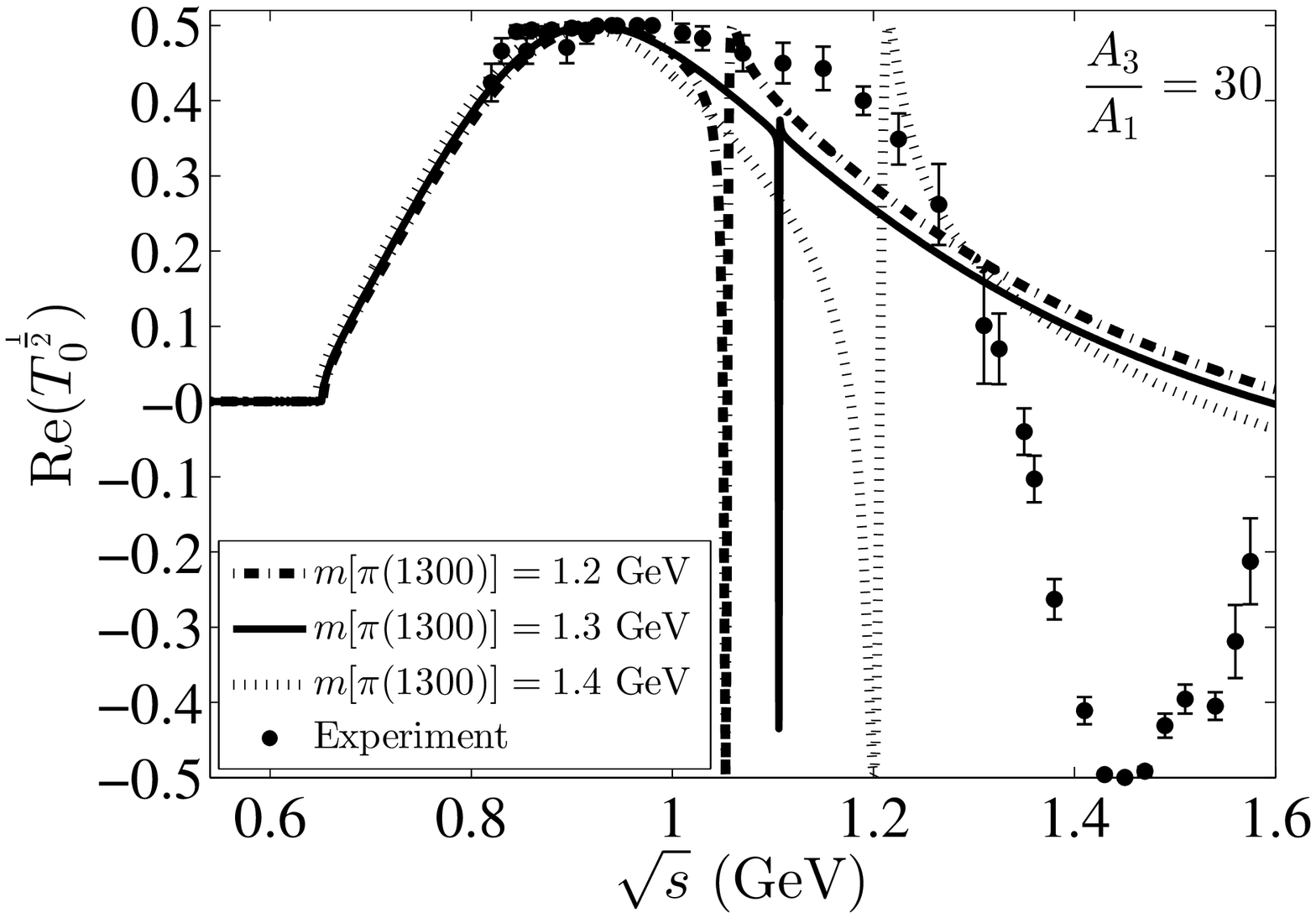}
 \caption{ Comparison of the generalized linear sigma model predictions for the real part of the $I=1/2$, $J=0$, $\pi K$ scattering amplitude with experimental data extracted from \cite{Aston}.   The predictions agree with data up to about 900 MeV range.}
 \label{F_ReT120_1.6GeV}
\end{center}
\end{figure}

\begin{figure}[!htb]
\begin{center}
\vskip 1cm
\epsfxsize = 10cm
 \includegraphics[width=15cm]{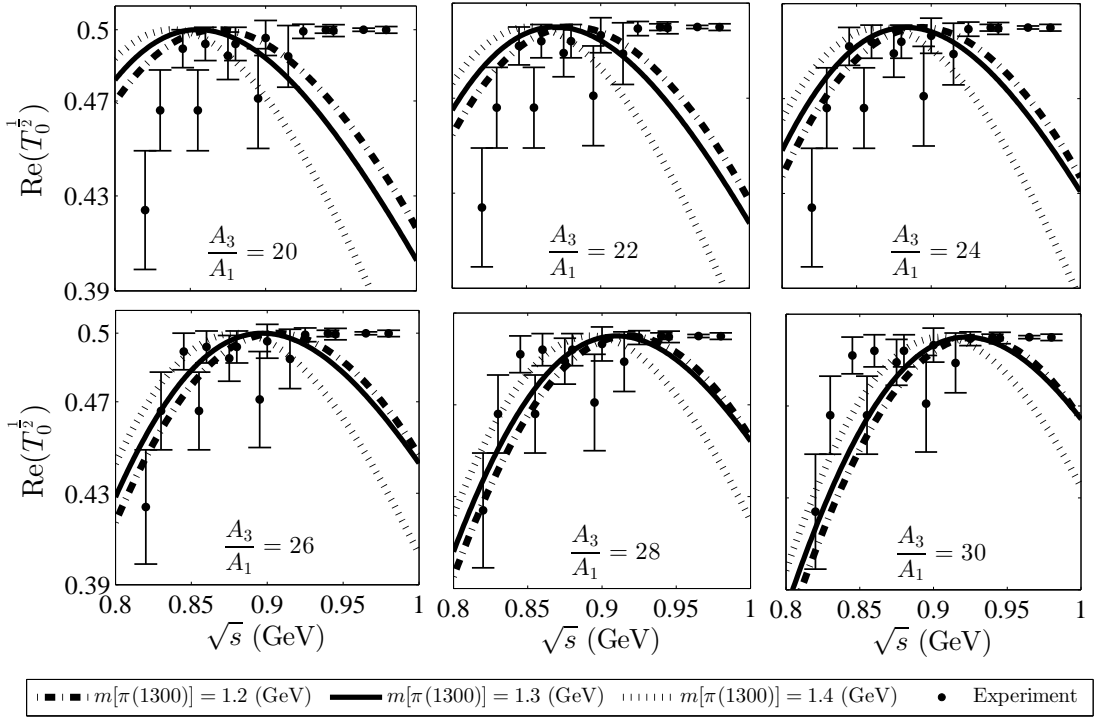}
\hskip 1cm
 \caption{ 
 	 Comparison of the generalized linear sigma model predictions in the low-energy region for the real part of the $I=1/2$, $J=0$, $\pi K$ scattering amplitude with experimental data extracted from \cite{Aston}.   The predictions agree with data up to about 900 MeV range.}
 \label{F_ReT120 1GeV}
\end{center}
\end{figure}

\begin{figure}[!htb]
\begin{center}
\vskip 1cm
\epsfxsize = 10cm
 \includegraphics[width=15cm]{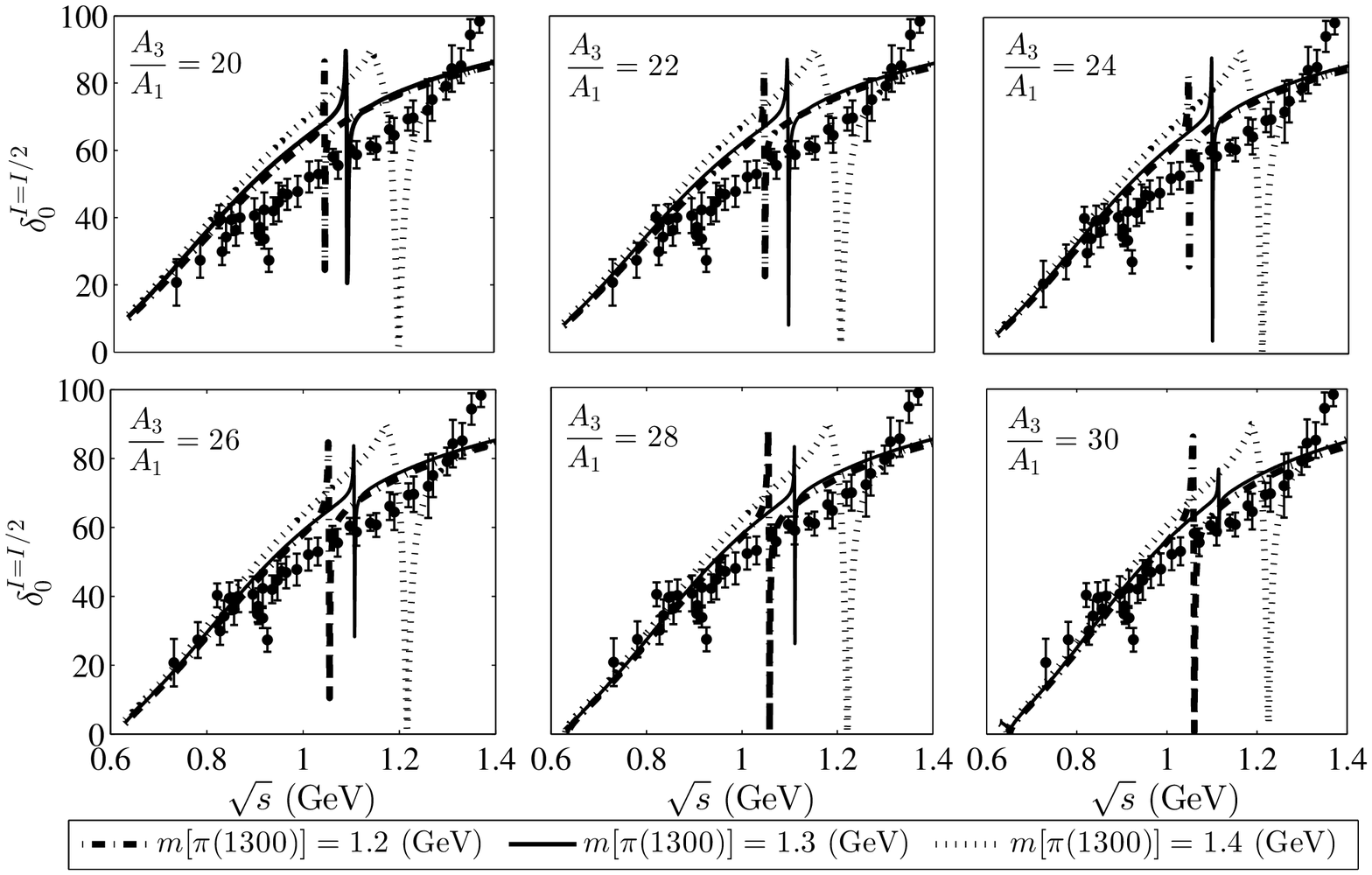}
\hskip 1cm
 \caption{Comparison of the generalized linear sigma model predictions for the phase shift in the  $I=1/2$, $J=0$ channel with experimental data of \cite{Aston}.   The predictions agree with data up to slightly above 1 GeV.}
 \label{ph_sh}
\end{center}
\end{figure}

For comparison with experiment it is convenient to focus on the
real part of the partial wave scattering amplitude in Eq. (\ref{T012_unitary}).
For typical values of the parameters we find the behavior of the bare amplitude (\ref{T012_B}) illustrated in Fig. \ref{F_generic_T120} (left).
As we discussed in (\ref{bare_amp_cond}) zeros that occur in the unitarized amplitude either result from
a zero or a pole  in $T_0^{\frac{1}{2}B}$. For comparison,  the same Fig. \ref{F_generic_T120}  shows both the ``bare'' amplitude (left) and the unitarized amplitude (right) in which circles and squares respectively show locations of zeros and poles in the ``bare'' amplitude.
We compare the model predictions for the scattering amplitude with the corresponding experimental data.      Although we do not expect the predictions to be accurate above around 1 GeV, nevertheless, we first plot the predicted amplitude up to 1.6 GeV in Fig.  \ref{F_ReT120_1.6GeV}.   We see that the amplitude is reasonably well predicted up to around 900 MeV.    Above this region,  the model requires inclusion of additional effects in the global analysis of Ref. \cite{global} which have been, for simplicity, neglected at the present level.     The additional effects include higher order terms in the chiral invariant part of the potential,  inclusion of higher order explicit SU(3) symmetry breaking, and inclusion of scalar and pseudoscalar glueballs (which do not directly affect the present channel but indirectly can have noticable effects through modification of the properties of isosinglets as well as some shifts in the Lagrangian parameters and couplings).
The figures show the variation of the amplitude with respect to the main uncertainties in the predictions that stem from two of the experimental inputs used to determine the parameters \cite{global}.   These are (a) the values of the SU(3) symmetry breaking
parameter $A_3/A_1$, and  (b) choices of the only roughly  known "heavy pion" mass $m[\Pi(1300)]$.
Similar to the case of $\pi\pi$ scattering amplitude studied in \cite{mixing_pipi}, we see that, without using any new parameters,
the mixing mechanism of \cite{global} predicts the scattering amplitude in reasonable  qualitative agreement with the low-energy experimental
data up to  around 900 MeV. This provides further  support for the validity of this mixing mechanism.
The predicted amplitude in the low-energy region is given in  Fig. \ref{F_ReT120 1GeV}.   Moreover, 
the model prediction for the $\pi K$ phase shift is compared with data in Fig. \ref{ph_sh} showing a close agreement up to slightly above 1 GeV.\\

\begin{figure}[!htb]
	\begin{center}
		\vskip 1cm
		\epsfxsize = 7.8cm
		\includegraphics[width=5 cm]{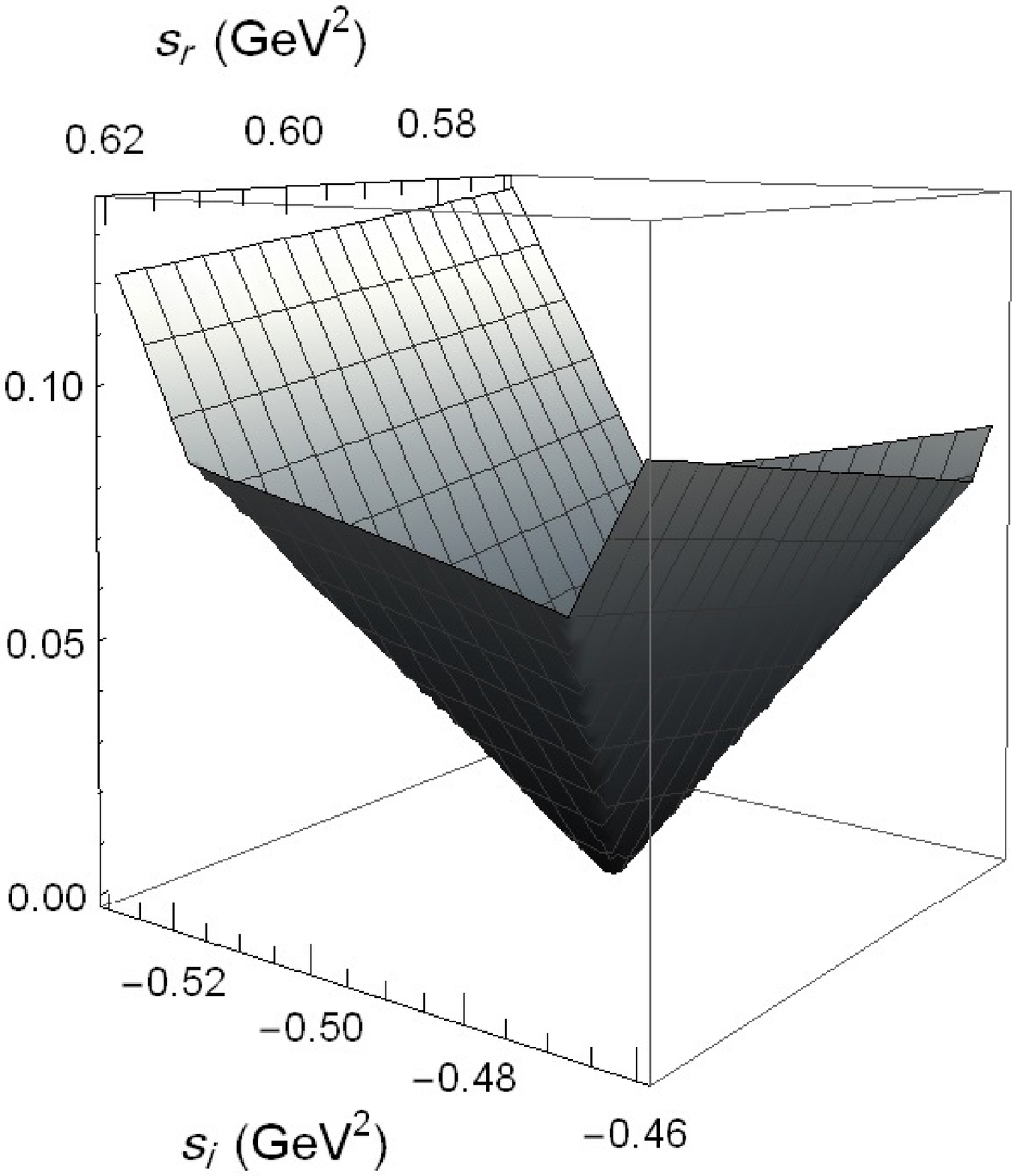}
		\hskip 1cm
		\epsfxsize = 7.5cm
		\includegraphics[width=5 cm]{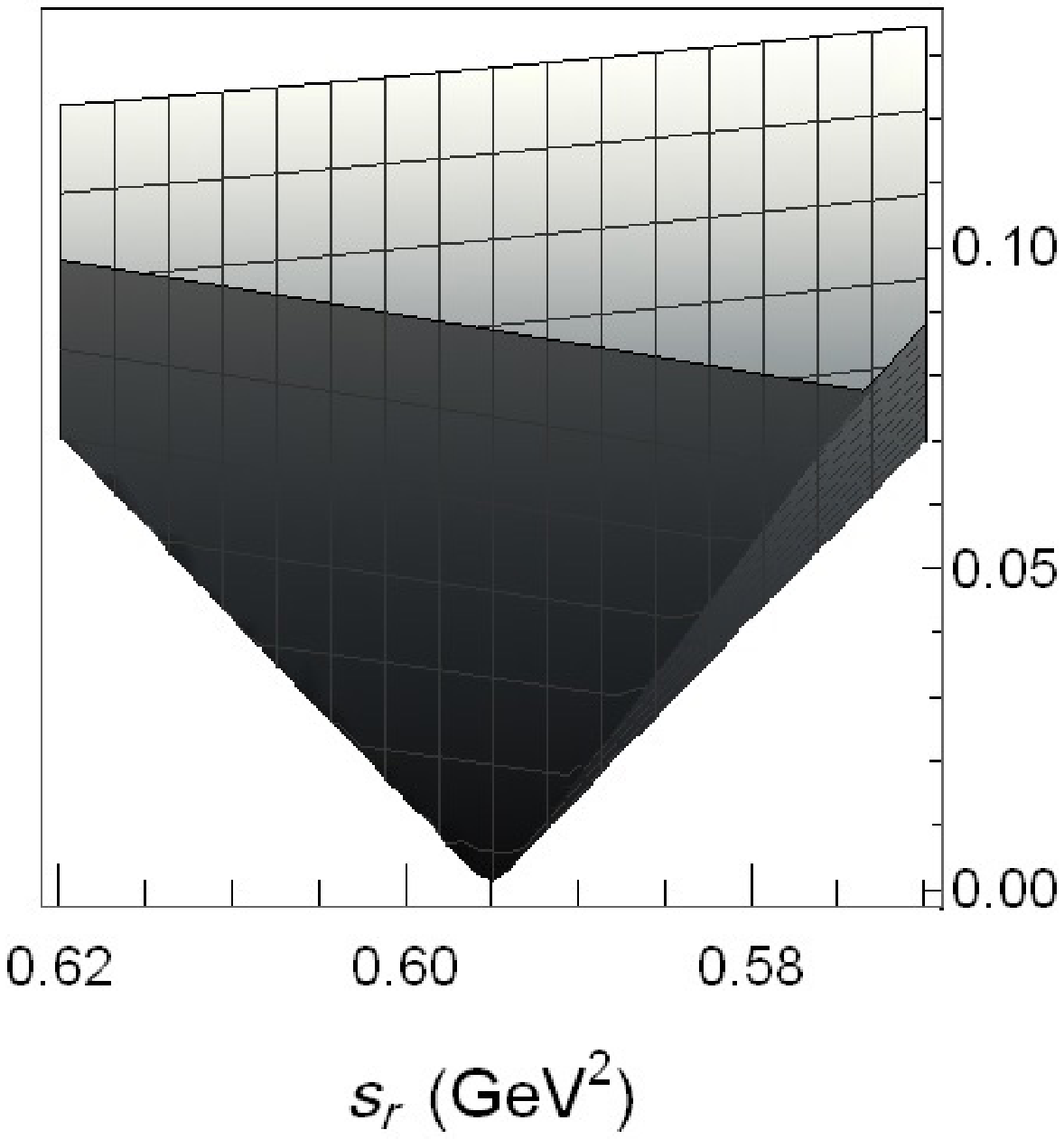}
		\hskip 0.2cm
		\epsfxsize = 7.5cm
		\includegraphics[width=5 cm]{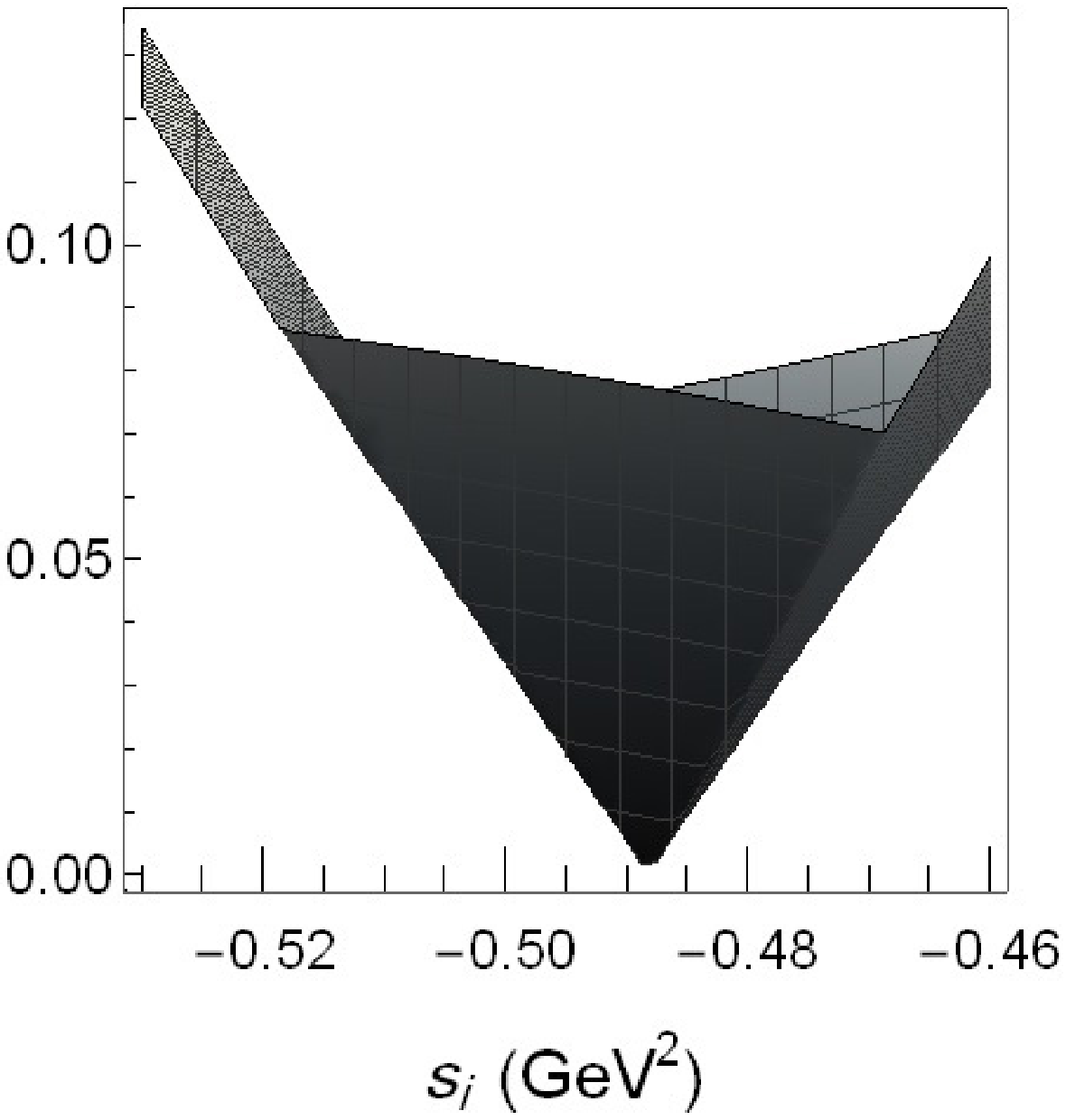}
		\vskip .1cm
		\epsfxsize = 7.8cm
		\includegraphics[width=5 cm]{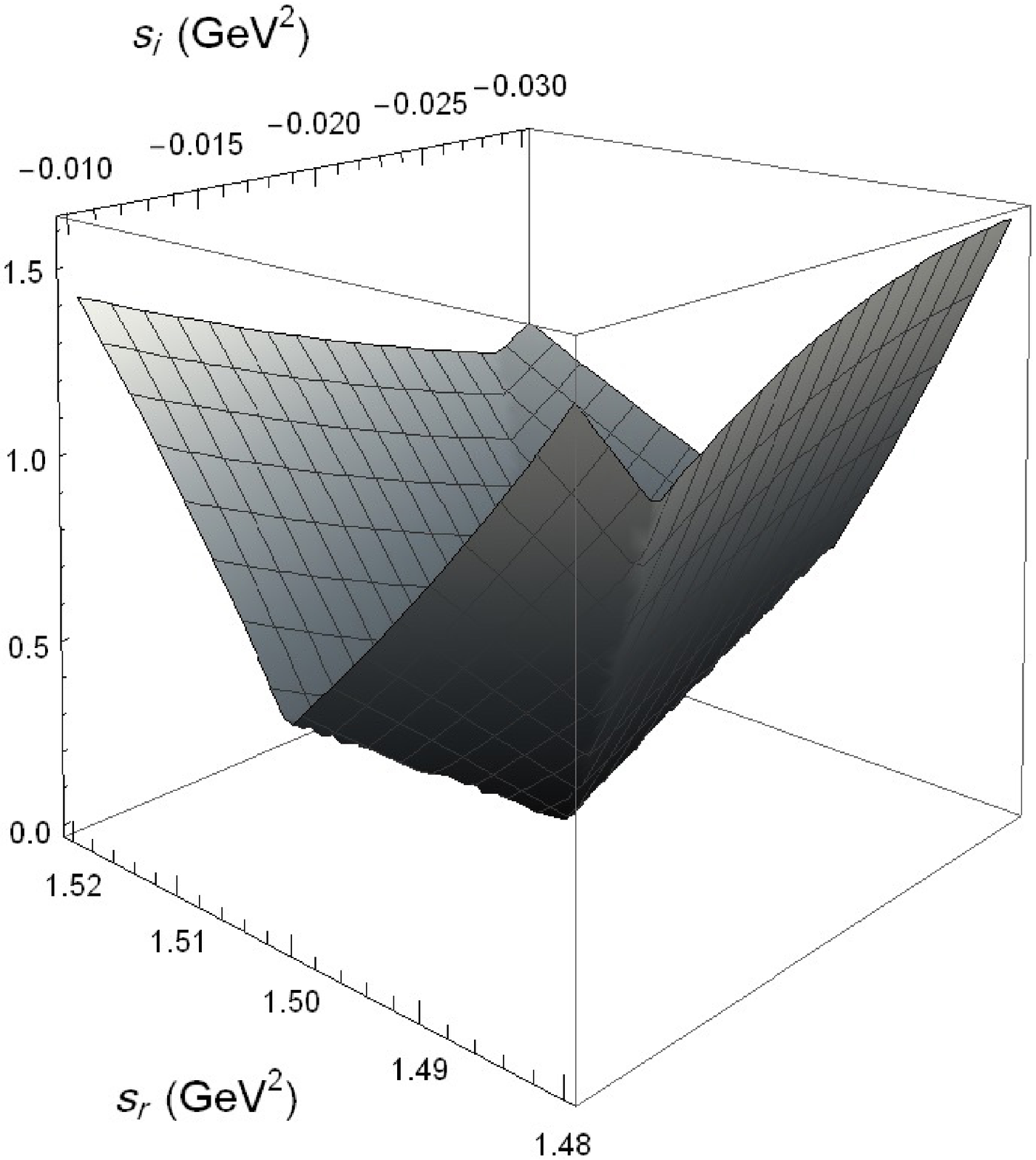}
		\hskip 1cm
		\epsfxsize = 7.5cm
		\includegraphics[width=5 cm]{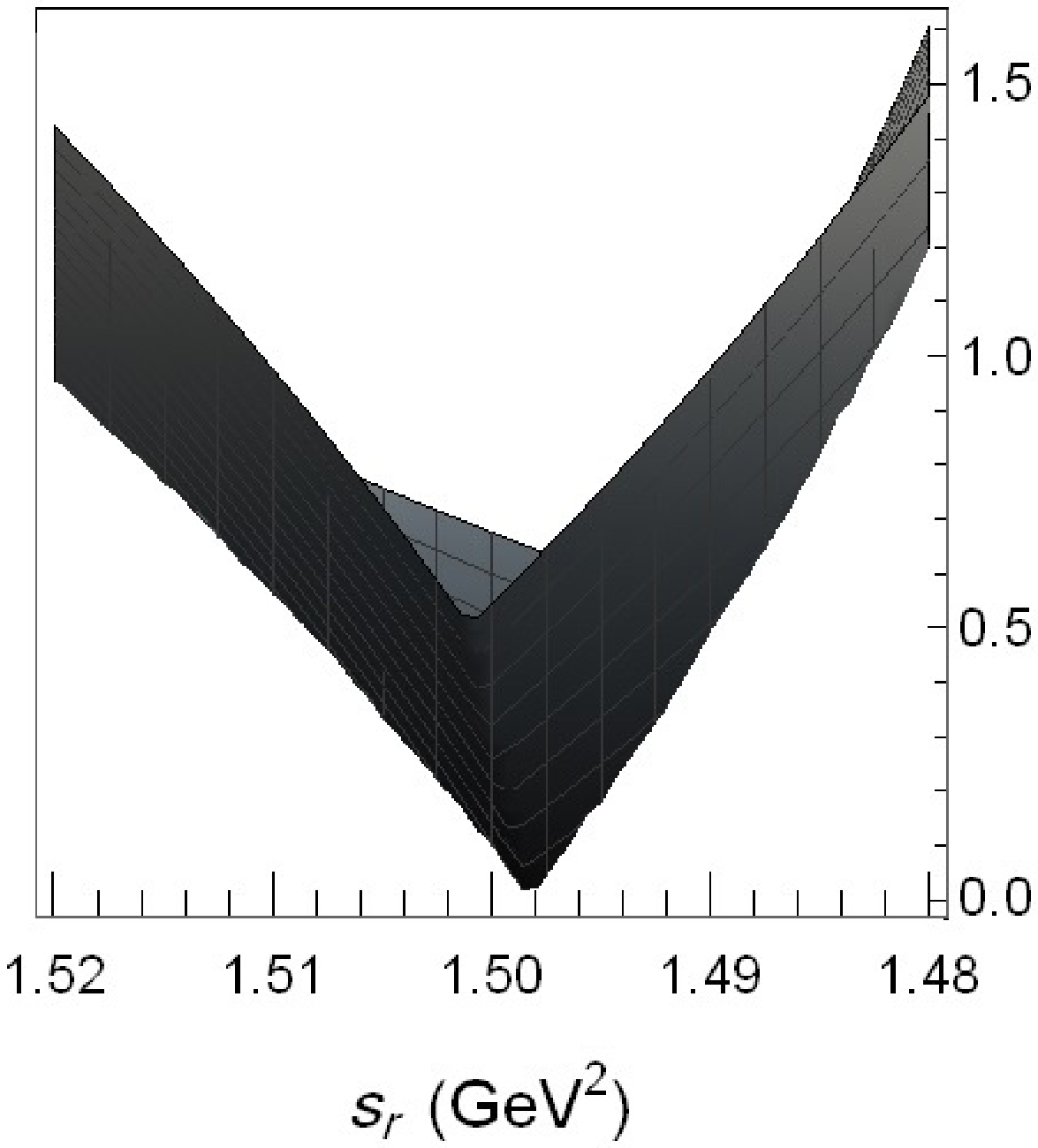}
		\hskip 0.2cm
		\epsfxsize = 7.5cm
		\includegraphics[width=5 cm]{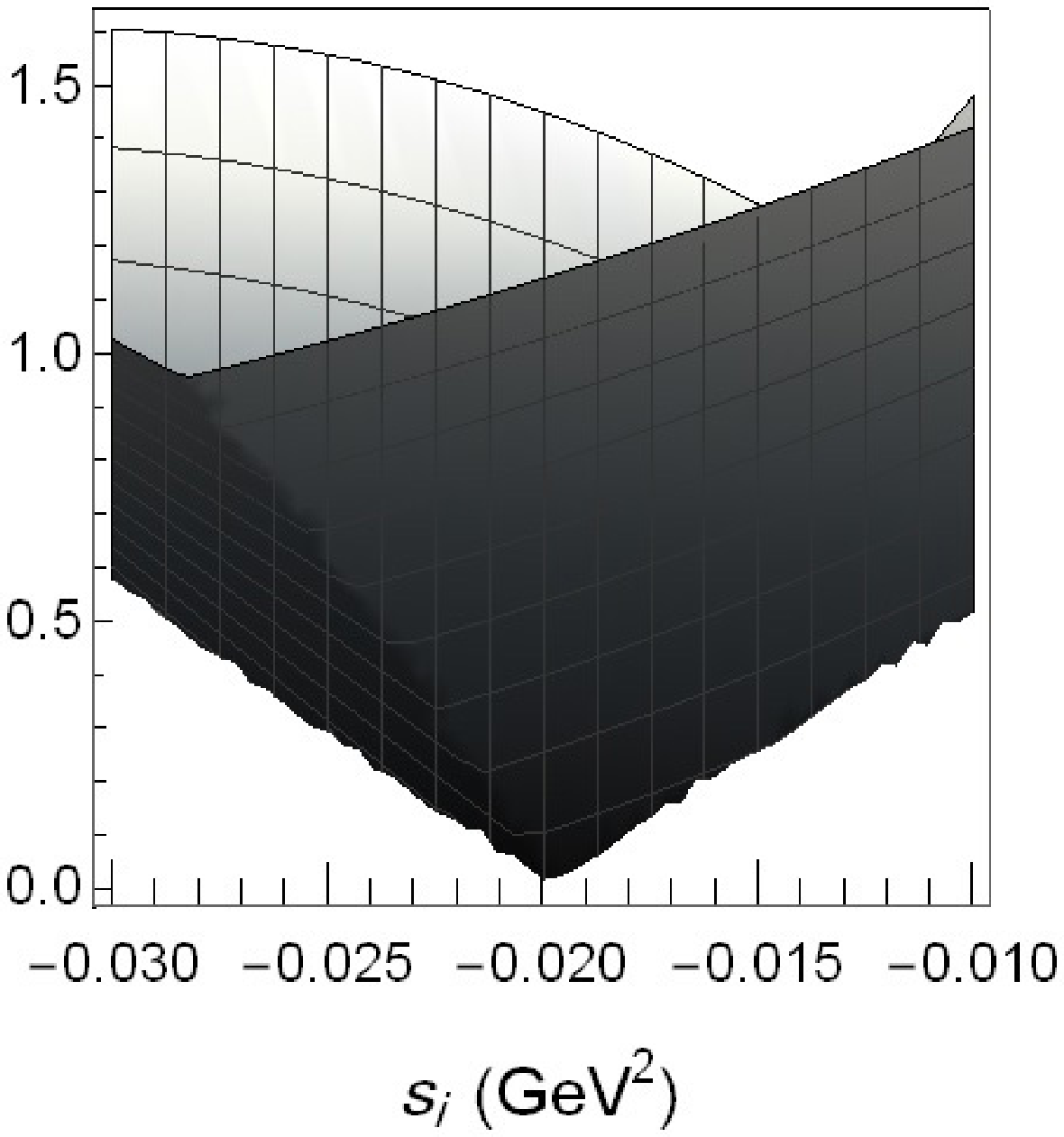}
		\caption{Plots of ${\cal F} \left(s_r, s_i\right)$ [defined in (\ref{F_srsi})] for $m[\pi(1300)]$ = 1.4 GeV and $A_3/A_1$=30.    The first (second) row shows the location of $\kappa_1$ ($\kappa_2$) pole.  In each row,  the three subplots from left to right respectively represent the 3D plot of  ${\cal F}$ and its projections onto  ${\cal F}-s_r$ and  ${\cal F}-s_i$ planes. }
		\label{3D_plot}
	\end{center}
\end{figure}

\begin{figure}[!htb]
	\begin{center}
		\vskip 1cm
		\epsfxsize = 7.5cm
		\includegraphics[width=8cm]{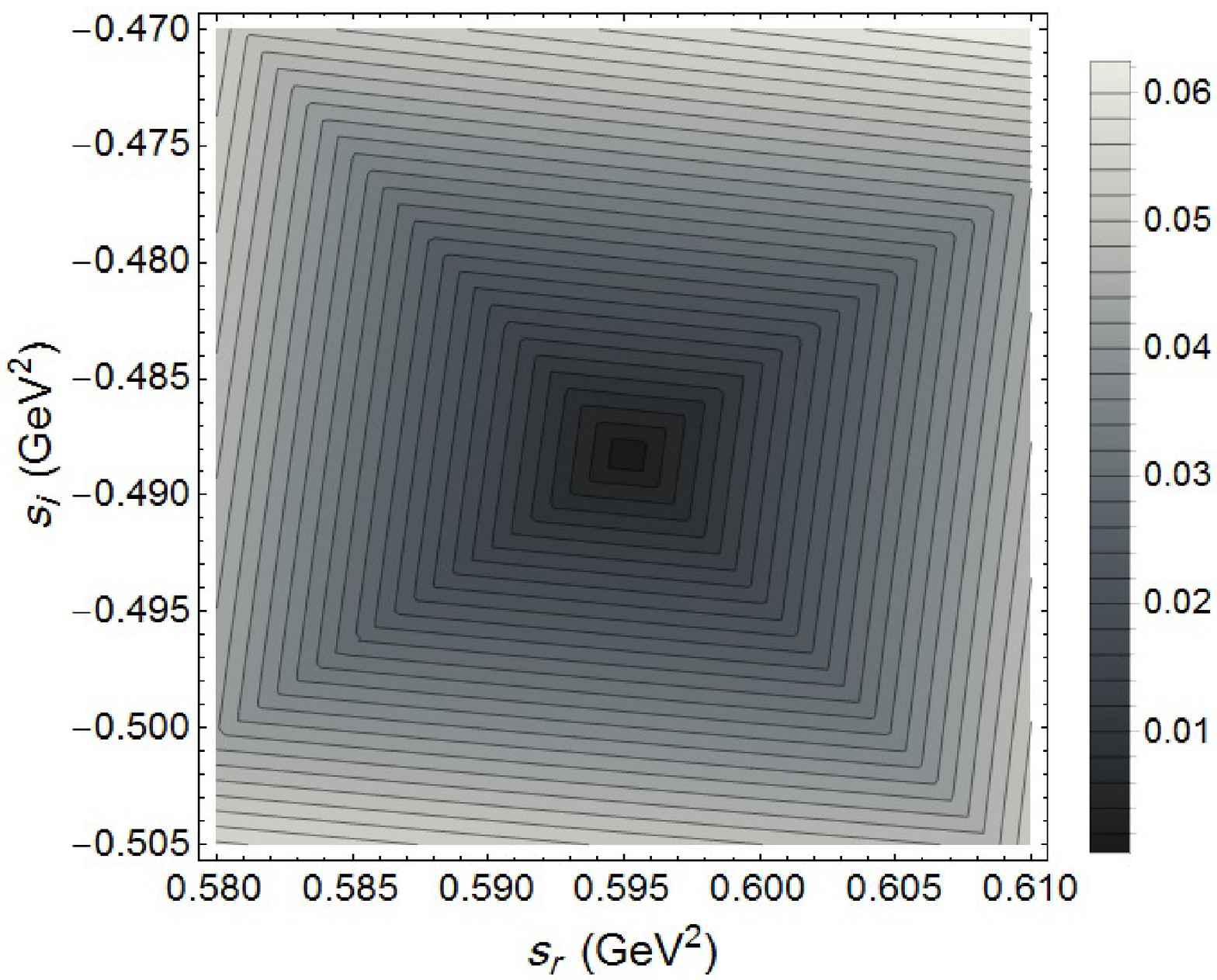}
		\hskip 1cm
		\epsfxsize = 7.5cm
		\includegraphics[width=8cm]{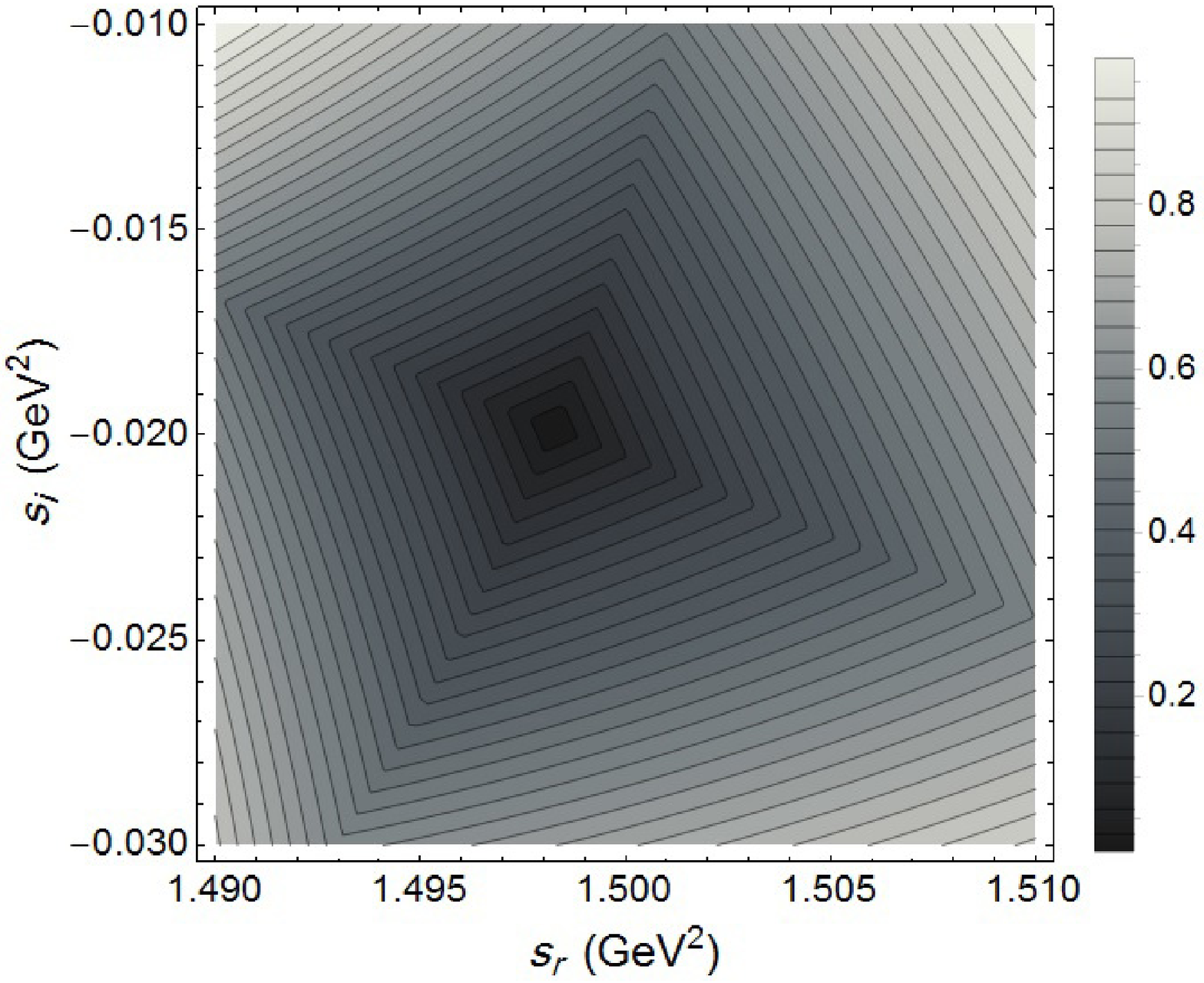}
		\caption{Contour plots of ${\cal F} \left(s_r, s_i\right)$ [defined in (\ref{F_srsi})] for $m[\pi(1300)]$ = 1.4 GeV and $A_3/A_1$=30.    Left (right) figure shows the location of $\kappa_1$  ($\kappa_2$) pole.}
		\label{con_plot}
	\end{center}
\end{figure}

\begin{figure}[!htb]
\begin{center}
\vskip 1cm
\epsfxsize = 9cm
 \includegraphics[width=11 cm]{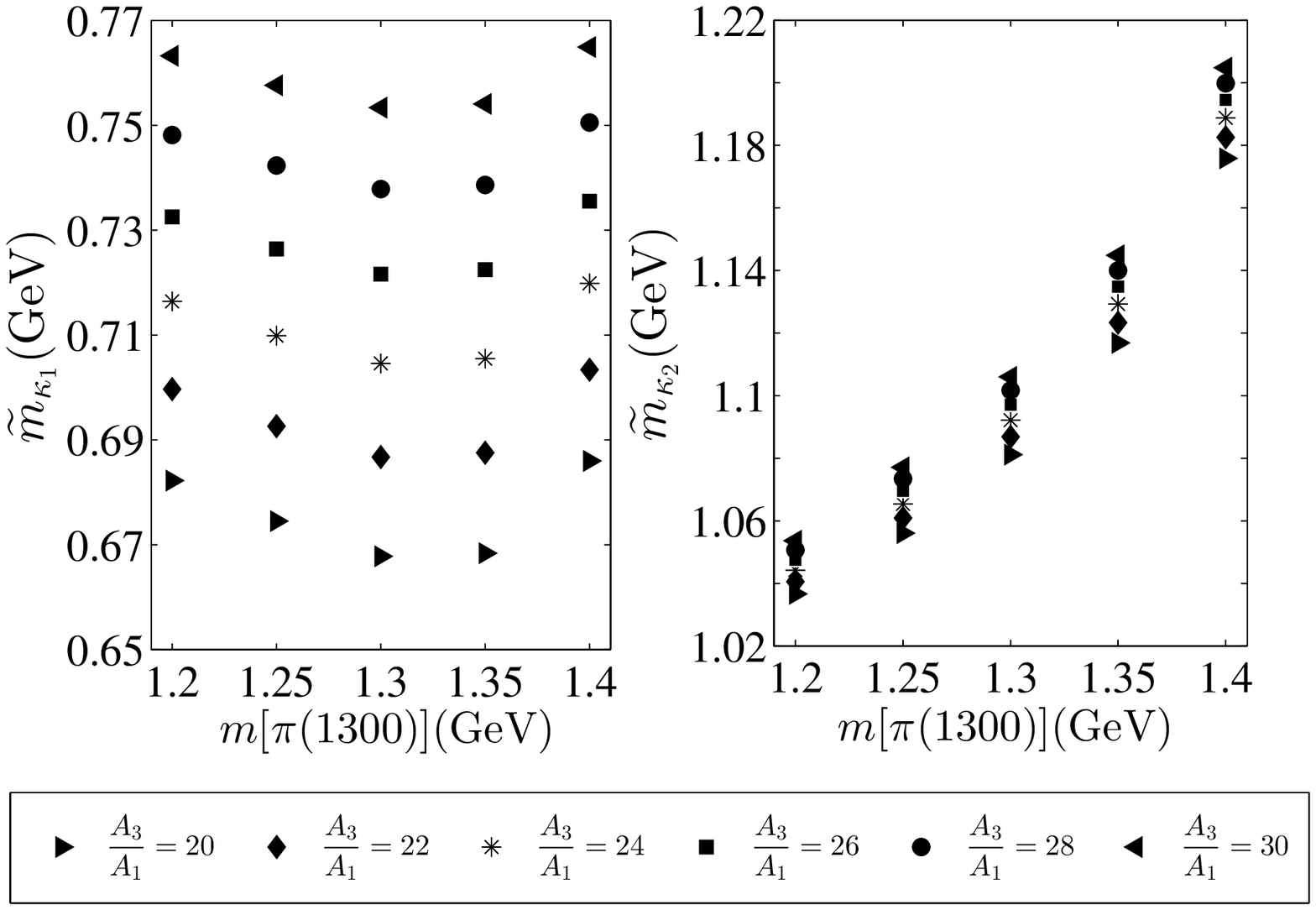}
 \caption{The dependencies of  physical masses of $\kappa_1$ and $\kappa_2$ mesons on $m[{\pi}(1300)]$ for different values of $A_3/A_1$.}
 \label{F_mass_phys}
\end{center}
\end{figure}

\begin{figure}[!htb]
\begin{center}
\vskip 1cm
\epsfxsize = 9cm
 \includegraphics[width=11 cm]{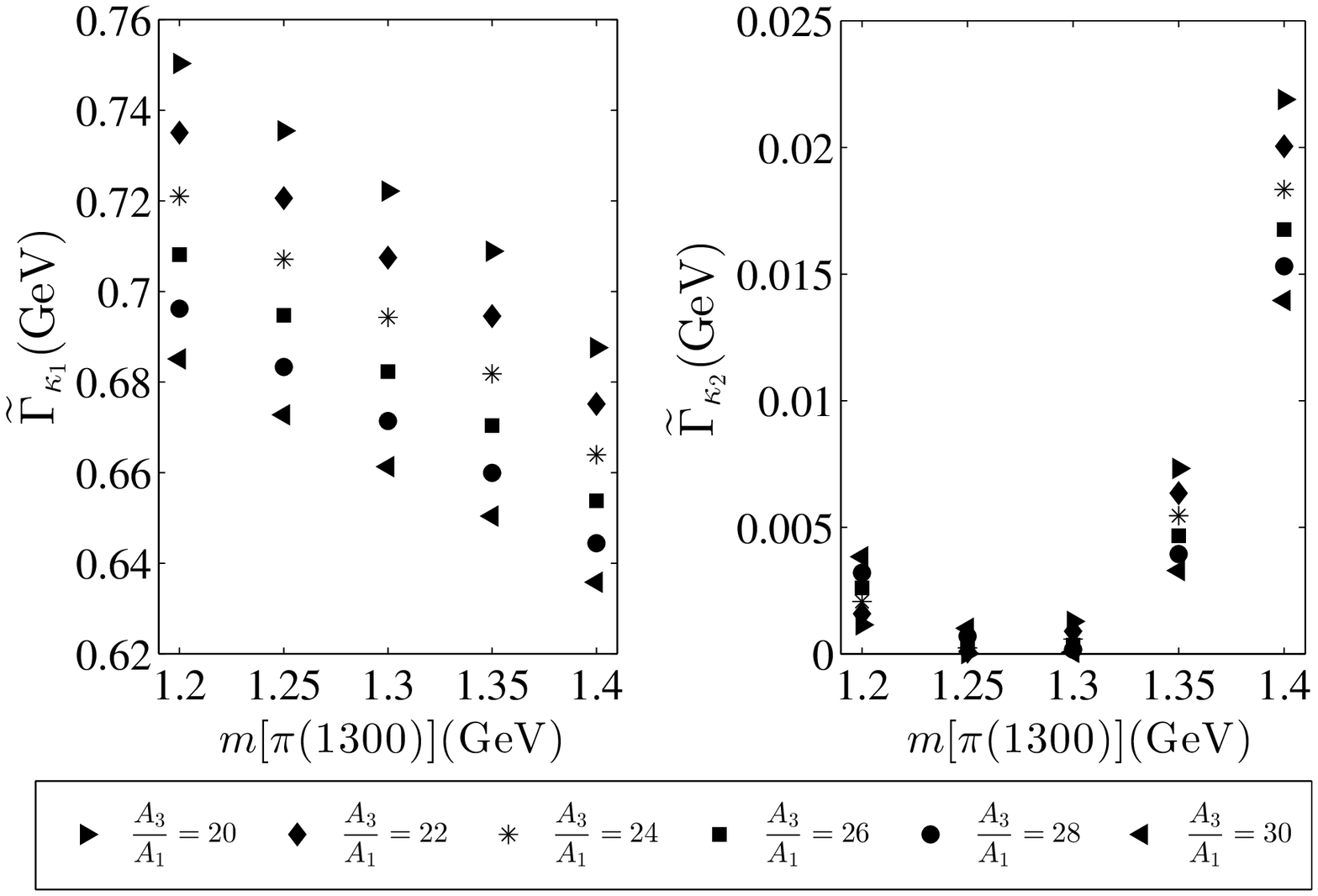}
\hskip .5cm
 \caption{The dependencies of physical widths of $\kappa_1$ and $\kappa_2$ mesons on $m[{\pi}(1300)]$ for different values of $A_3/A_1$.}
\label{F_decay_phys}
\end{center}
\end{figure}

\begin{figure}[!htb]
\begin{center}
\vskip 1cm
\epsfxsize = 9cm
 \includegraphics[width=11 cm]{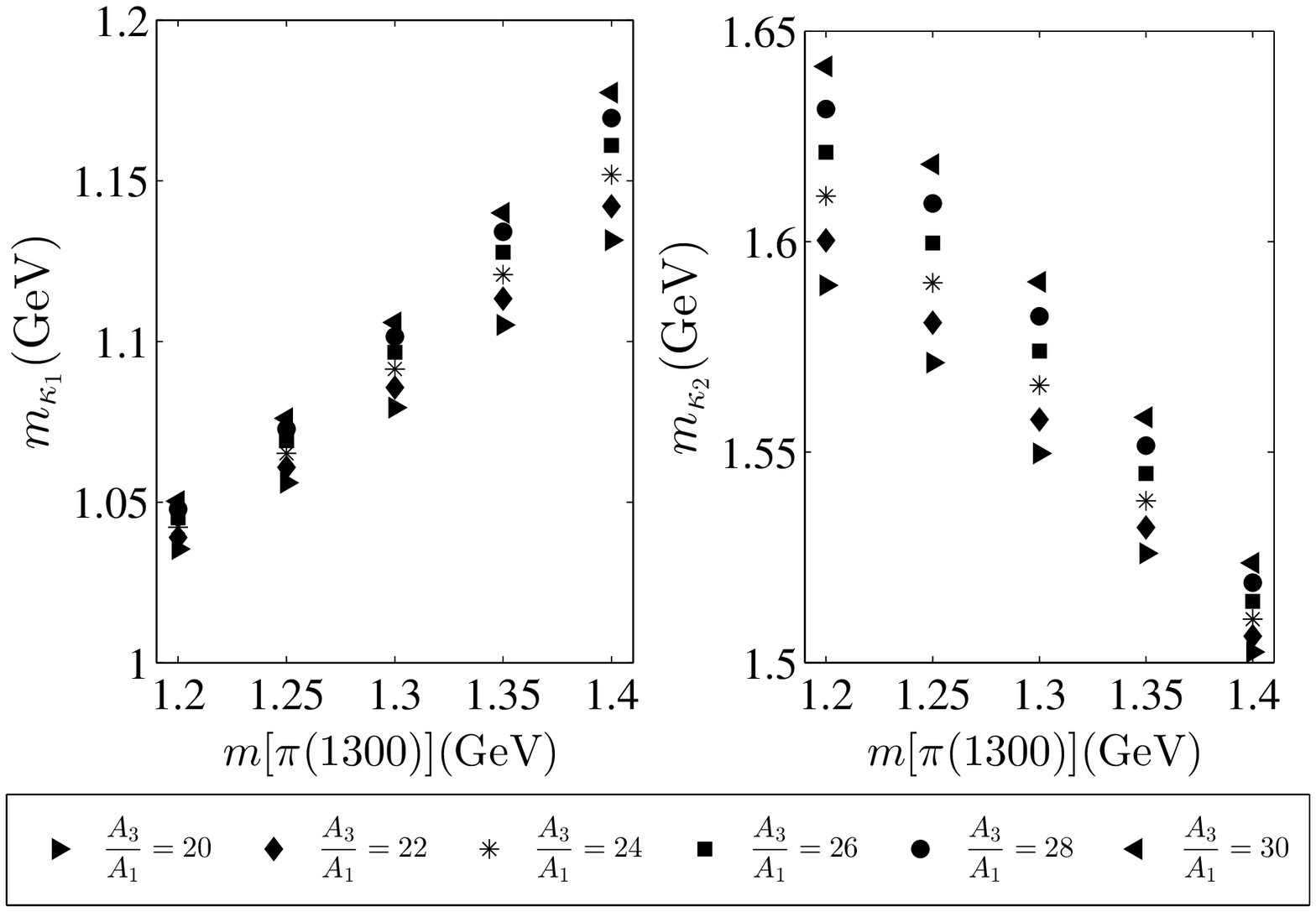}
 \caption{The dependencies of  bare masses of $\kappa_1$ and $\kappa_2$ mesons on $m[{\pi}(1300)]$ for different values of $A_3/A_1$.}
 \label{F_mass_bare}
\end{center}
\end{figure}

Next, we examine the poles of the K-matrix unitarized amplitude.    In the case of $\pi\pi$ scattering \cite{mixing_pipi} the first pole clearly captured the properties of light and broad sigma and the second pole resembled the $f_0(980)$.
For interpretation of the physical resonances it is
conventional to look at the pole positions in the complex plane of the
analytically continued expression for $T_0^{\frac{1}{2}}$.
We examine these physical pole positions by
solving for the complex roots of the
denominator of the  K-matrix unitarized amplitude
Eq. (\ref{T012_unitary}):
\begin{equation}
{\cal D} (s) = 1 - i\, T_0^{\frac{1}{2}B} = 0,
\label{pole_eq}
\end{equation}
with $(T_0^{\frac{1}{2}})^B$ given by Eq. (\ref{T012_B}).   We
search for solutions, $s^{(j)}= s_r^{(j)}+ i s_i^{(j)} = {\widetilde m}_j^2 - i {\widetilde m}_j {\widetilde\Gamma_j}$ of this equation, where
${\widetilde m}_j$ and ${\widetilde \Gamma}_j$ are interpreted as the
physical mass and decay width of the $j$-th resonance. A
first natural attempt would be to simultaneously solve the two equations:
\begin{eqnarray}
{\rm Re}{\cal D} \left(s_r, s_i\right) = 0, \nonumber \\
{\rm Im}{\cal D} \left(s_r, s_i\right) = 0.
\label{ReandIm}
\end{eqnarray}
However, this approach turns out to be rather tedious to be  implemented. A more efficient numerical approach, that was  first presented in \cite{mixing_pipi},  is to consider the positive function
\begin{equation}
{\cal F} \left(s_r, s_i\right) =
\left| {\rm Re}
\left(
{\cal D} (s_r, s_i)
\right)\right| +
\left| {\rm Im}
\left(
{\cal D} (s_r, s_i)
\right)\right|,
\label{F_srsi}
\end{equation}
which allows determination of poles by searching for the zeros of this function.    Fig. \ref{3D_plot} shows the 3D plot of
${\cal F} \left(s_r, s_i\right)$ in the neighborhood of the two isodoublet poles  that the model predicts ($\kappa_1$ and $\kappa_2$), 
and Fig. \ref{con_plot} gives the contour plot of this function  over the complex $s$-plane for values of $m[\pi(1300)]$=1.4 GeV and $A_3/A_1$=30. 
We further solve for the exact location of the first and the second poles which consequently yield the physical masses and the decay widths.   These are displayed in Figs. \ref{F_mass_phys} and \ref{F_decay_phys} versus $m[\Pi(1300)]$ for several values of $A_3/A_1$.   It is evident that the  model predicts a light and broad isodoublet scalar meson in the $\pi K$ channel in complete parallel
to its prediction of a light and broad sigma in the $\pi\pi$ channel.    It is seen that the mass of $\kappa$ meson [or $K^*_0(800)$ in the PDG listing \cite{pdg}] is predicted to be in the range of 670-770 MeV and its decay width in the range of 640-750 MeV. This provides further support for the  generalized linear sigma model \cite{global} and the global picture of scalar and pseudoscalar mesons below and above 1 GeV.   The mass of the second isodoublet state also receives considerable unitarity corrections, but since the states above 1 GeV require additional effects that have been ignored here for simplicity, we do not further analyze it here.    For comparison, the ``bare'' masses are given in Fig. \ref{F_mass_bare}.

\subsection{Pole expansion}
\begin{figure}[!htb]
\centering
     \includegraphics[height=2.5in]{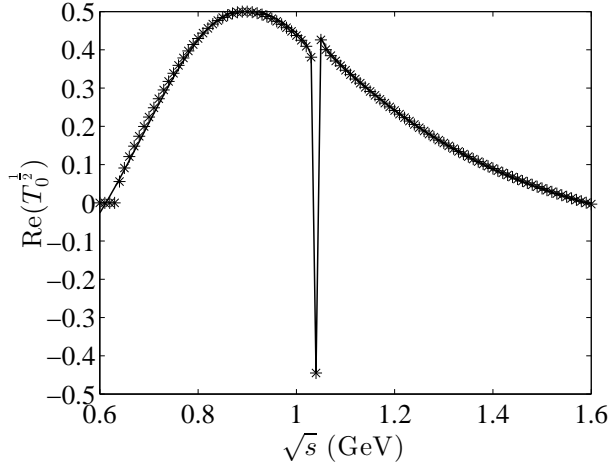}
                        \caption{Comparing the K-matrix unitarized $I=1/2$, $J=0$, $\pi K$ scattering amplitude with the expansion (\ref{pole_expan}) for $ A_3/A_1=20 $ and $ m[{\pi}(1300)]=1.2$.}
\label{F_I12_compare}
\end{figure}

In this subsection we highlight an interesting property of the unitarization methodology applied in this work.     Organizing the bare amplitude  in terms of the poles and a remaining background
\begin{eqnarray}
T_0^{\frac{1}{2}B}&=&T_{\alpha}+\sum_{i=1}^{n_{\kappa}}{\frac{T_{\beta}^i}{m_{\kappa_i}^2-s}} = \frac{\rho (s)}{2} \left[
T'_{\alpha} + \sum_{i=1}^{n_{\kappa}}{\frac{{T'}_{\beta}^i}{m_{\kappa_i}^2-s}}\right],
\end{eqnarray}
where
\begin{eqnarray}
T'_{\alpha}&=&-2\gamma_{\pi K}^{(4)}-\frac{1}{4q^2}\sum_{i=1}^{n_{\kappa}} {\gamma_{\kappa_i\pi K}^2}\ln{\left(\frac{B_{\kappa_i}+1}{B_{\kappa_i}-1}\right)}+\frac{1}{2q^2}\sum_{j=1}^{n_f} \gamma_{f_{j}KK}\gamma_{f_{j}\pi\pi}\ln{\left(1+\frac{4q^2}{m_{f_j}^2}\right)},  \\
{T'}_{\beta}^i&=&3 \gamma_{\kappa_i \pi K}^2,
\label{T012_alpha_beta}
\end{eqnarray}
we can show that the K-matrix unitarized amplitude has a similar mathematical structure (in the complex plane) and can be written as a sum  of complex poles and a constant complex background
\begin{eqnarray}
T_0^{\frac{1}{2}}=\frac{T_0^{\frac{1}{2}B}}{1-i T_0^{\frac{1}{2}B}}
\approx
\widetilde{T}_{\alpha}+\sum_{i=1}^{n_{\kappa}}{\frac{\widetilde{T}_{\beta}^i}{\widetilde{m}_{\kappa_i}^2-s-i\widetilde{m}_{\kappa_i}\widetilde{\Gamma}_{\kappa_i}}}
\approx
\frac{\rho (s)}{2}
\left[
\widetilde{T}'_{\alpha}+\sum_{i=1}^{n_{\kappa}}{\frac{\widetilde{T}_{\beta}^{\prime i}}{\widetilde{m}_{\kappa_i}^2-s-i\widetilde{m}_{\kappa_i}\widetilde{\Gamma}_{\kappa_i}}}
\right]
\label{pole_expan}
\end{eqnarray}
which shows that the functional form of the K-matrix unitarized amplitde resembles the bare amplitude in which the bare masses are replaced by the physical poles in the complex $s$-plane.
The real part of the $I=1/2$, $J=0$ scattering amplitude obtained from the expansion (\ref{pole_expan}) is verified numerically in Fig. \ref{F_I12_compare}.

 Moreover,  the bare decay width  and mass of kappa's satisfy
\begin{equation}
m_{\kappa i} \Gamma_{\kappa_i} =
\left. \frac{\rho (s)}{2} {T}_{\beta}^{'i}\right|_{s=m_{\kappa_i}^2}
\end{equation}
which is again in parallel with the physical decay width and mass of kappa's
 \begin{equation}
 \widetilde{m}_{\kappa i} \widetilde{\Gamma}_{\kappa_i} \approx \left|\widetilde{T}_{\beta}^i\right| \approx
 \left. \frac{\rho (s)}{2} \widetilde{T}^{'i}_{\beta}\right|_{s=\widetilde{m}_{\kappa_i}^2 - i \widetilde{m}_\kappa \widetilde{\Gamma}_\kappa}
\label{decay_width_compare}
\end{equation}
This relationship is  numerically tested for two values of $A_3/A_1$ over the range of $m[\pi(1300)]$ in Fig. \ref{decay_width_compare}.

\begin{figure}[!htb]
\begin{center}
\vskip 1cm
\epsfxsize = 9cm
 \includegraphics[width=10cm]{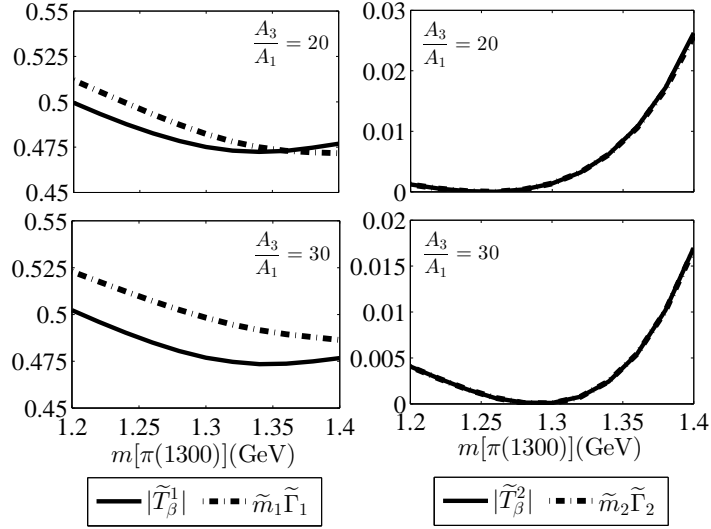}
 \caption{Comparison of $\widetilde{m}_{\kappa_i}\widetilde{\Gamma}_{\kappa_i}$ (dot-dashed line) with $\left|\tilde{T}_\beta\right|$ (solid line).}
 \label{decay_width_compare}
\end{center}
\end{figure}

\subsection{$I=3/2$ Results}

\begin{figure}[!htb]
\begin{center}
\vskip 1cm
\epsfxsize = 10cm
 \includegraphics[width=15cm]{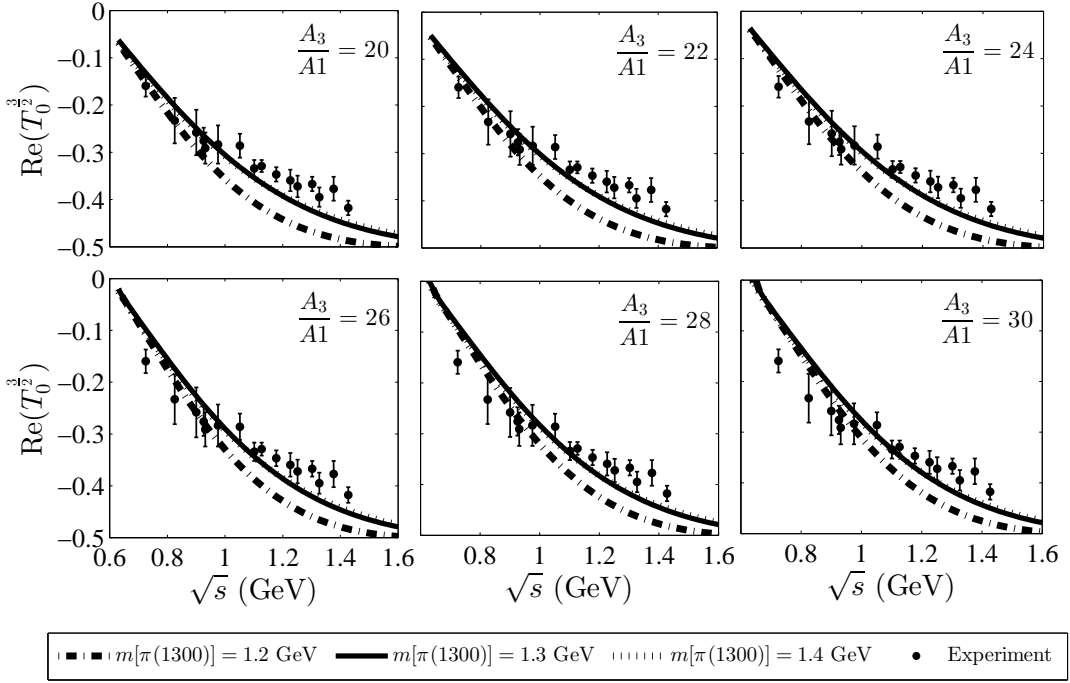}
\hskip 1cm
 \caption{Prediction of the generalzied linear sigma model for the $I=3/2$, $J=0$, $\pi K$ scattering amplitude  with experimental data.}
 \label{F_I3/2}
\end{center}
\end{figure}

We saw in Sec. III that the $I=1/2$, $J=0$,  $\pi K$ scattering amplitude can be fitted with various mathematical structures some of which as simple as a constant background and a pole.   The same can be said about the $I=3/2$, $J=0$ amplitude which can be independently fitted.  However, despite the success of these fits separately for each channel,   we saw that it is non-trivial to fit both at the same time.   In this section we present the prediction of the generalized linear sigma model for the $I=3/2$ channel with the same parameters used in the $I=1/2$ case. Fig. \ref{F_I3/2} compares these prediction for different values of $A_3/A_1$ and $m[\pi(1300)]$ with experimental data, showing a close   agreement up to about 1 GeV (and slightly thereafter).   We interpret the simultaneous agreement of the model predictions for both the $I=1/2$ and the $I=3/2$ channels (up to 1 GeV) as further support for the generalized linear sigma framework of Ref. \cite{global}.

 \subsection{Scattering Lengths}
Although the present framework is not specifically designed to probe the near threshold dynamics, we examine its predictions for the scattering lengths to further test the model.   Expanding the $S$-wave scattering amplitude near the threshold the scattering lengths are commonly defined \cite{bijnens_04}
\begin{equation}\label{scatlengthexpan}
t_0^{I\, B}=\frac{\sqrt{s}}{2}  (a_0^I+\frac{q^2}{m_\pi^2}b_0^I + \frac{q^4}{m_\pi^4}c_0^I+\cdots), \quad {\rm for} \: q\rightarrow 0,\, s\rightarrow(m_\pi+m_K)^2,
\end{equation}
where the partial wave amplitude $t_0^{I\, B}$ is related to our $T_0^{I\, B}$ defined in (\ref{T012B_def})
\begin{equation}
t_0^{I\, B}=\frac{\sqrt{s}}{2}T_0^{I\, B}=\frac{1}{32\pi}\int_{-1}^{1} d\cos\theta P_0(\cos\theta)A^I(s,t,u).
\end{equation}
Therefore
\begin{equation}
T_0^{I\, B}= q (a_0^I+\frac{q^2}{m_\pi^2}b_0^I + \frac{q^4}{m_\pi^4}c_0^I+\cdots), \quad {\rm for} \: q\rightarrow 0,\, s\rightarrow(m_\pi+m_K)^2.
\end{equation}
In the units of $({\rm pion\, scattering\, wavelength)}^{2l+1}$, the S-wave scattering lengths are then computed
\begin{eqnarray}
a_0^I &=& m_\pi \lim_{q\rightarrow 0} \frac{2}{\sqrt{s}}\, t_0^{I\, B}= m_\pi \lim_{q\rightarrow 0} \frac{1}{q}\, T_0^{I\, B},\nonumber\\
b_0^I &=&\frac{ m_\pi^3}{2!} \lim_{q\rightarrow 0} \frac{2}{\sqrt{s}}\, \frac{\partial^2 t_0^{I\, B}}{\partial q^2}=\frac{ m_\pi^3}{2!} \lim_{q\rightarrow 0} \frac{1}{q}\, \frac{\partial^2 T_0^{I\, B}}{\partial q^2},\nonumber\\
c_0^I &=& \frac{m_\pi^5}{4!} \lim_{q\rightarrow 0} \frac{2}{\sqrt{s}}\, \frac{\partial^4 T_0^{I\, B}}{\partial q^4}= \frac{m_\pi^5}{4!} \lim_{q\rightarrow 0} \frac{1}{q}\, \frac{\partial^4 T_0^{I\, B}}{\partial q^4}.
\end{eqnarray}

\begin{figure}[!htb]
\begin{center}
\vskip 1cm
\epsfxsize = 9cm
 \includegraphics[width=15 cm]{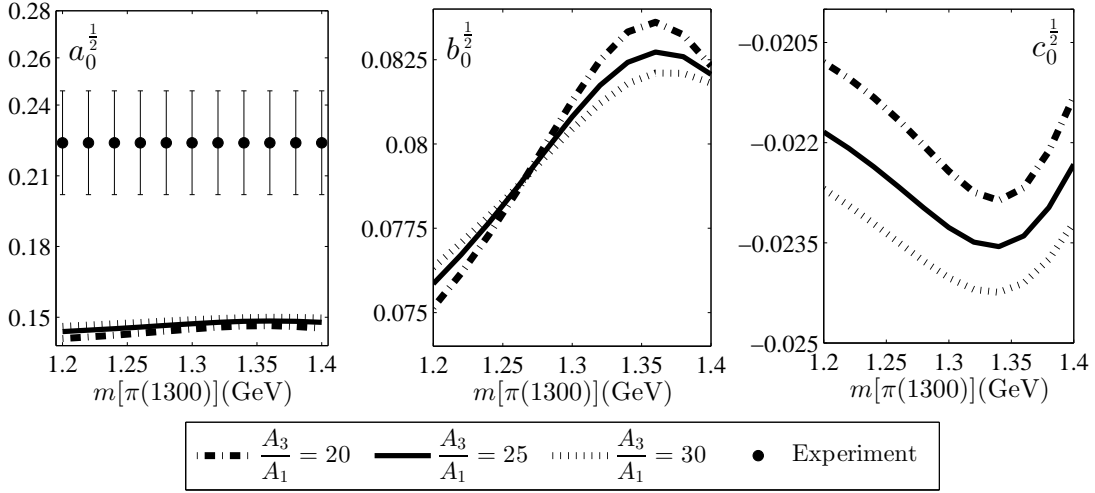}
 \caption{$\pi K $ scattering lengths computed with the bare scattering amplitude $T_0^{\frac{1}{2}B} $.}
 \label{F_SL_12}
\end{center}
\end{figure}

\begin{figure}[!htb]
\begin{center}
\vskip 1cm
\epsfxsize = 9cm
 \includegraphics[width=15 cm]{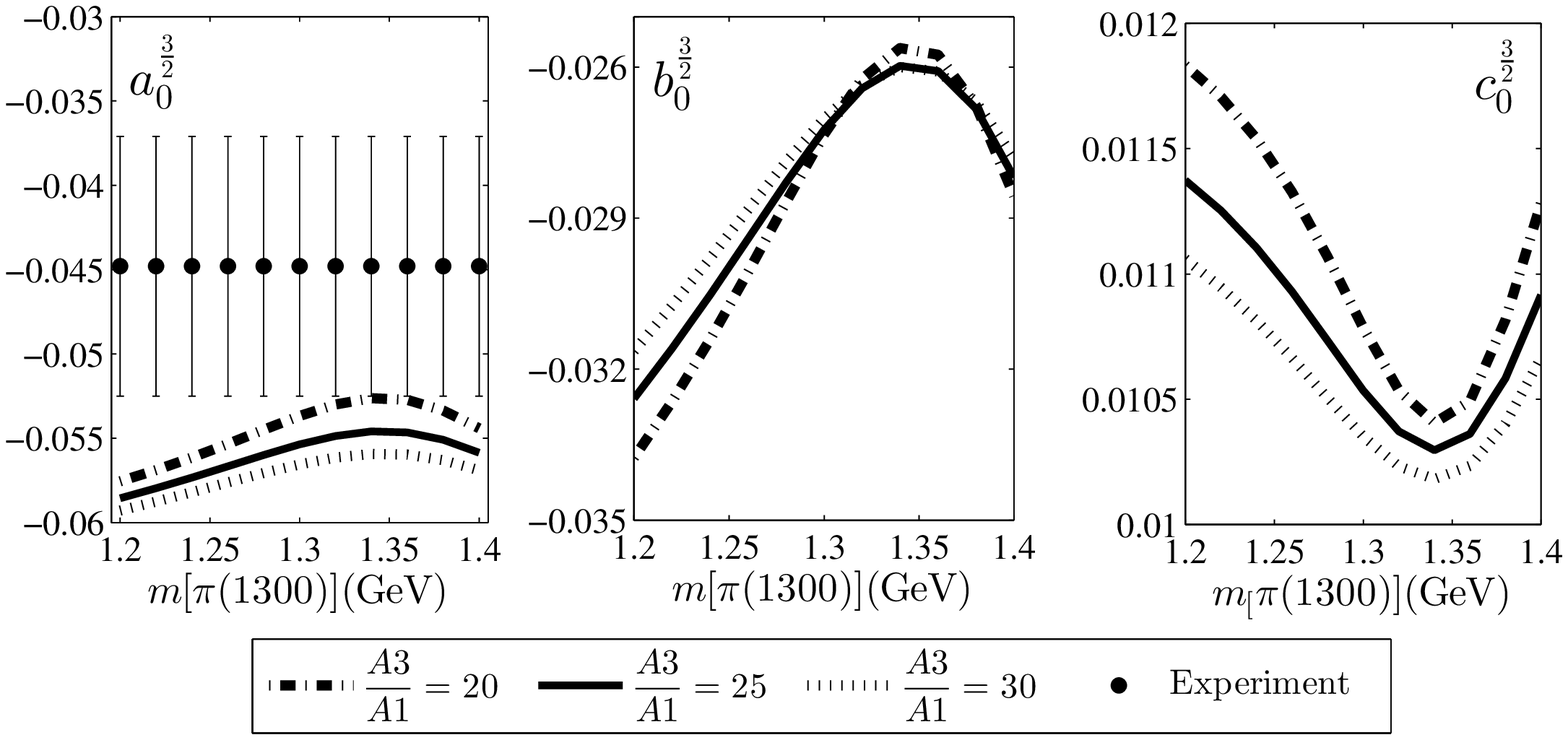}
 \caption{$\pi K $ scattering lengths computed with the bare scattering amplitude $T_0^{\frac{3}{2}B} $.}
 \label{F_SL_32}
\end{center}
\end{figure}

\begin{figure}[!htb]
\begin{center}
\vskip 1cm
\epsfxsize = 9cm
 \includegraphics[width=15 cm]{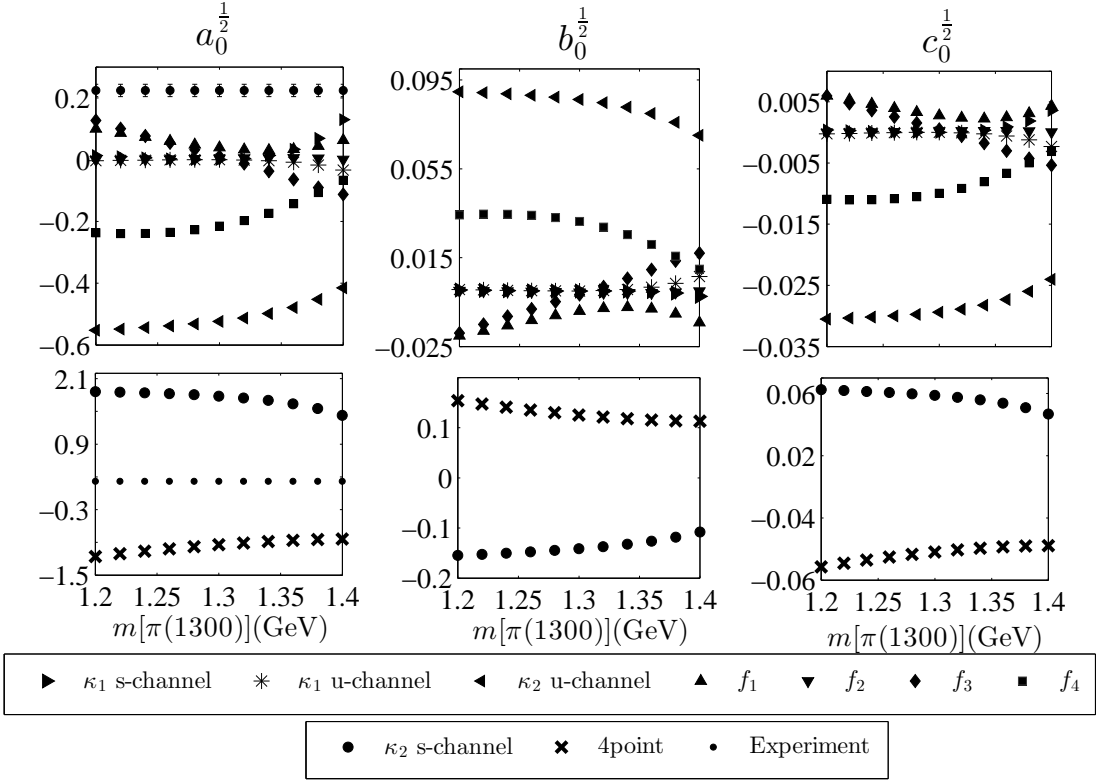}
 \caption{Individual contributions to $\pi K $ scattering lengths for $\frac{A_3}{A_1}=30$ using bare scattering amplitude $T_0^{\frac{1}{2}B}$.}
 \label{F_SL_12_IC}
\end{center}
\end{figure}

In the present framework,  these coefficients are not expected to be very accurate since the model is designed to give a global understanding of scalar and pseudoscalar mesons below and above 1 GeV and hence it spans a wide range of energy as opposed to zoom in on the near threshold region.    Nevertheless,  we have computed the scattering lengths computed in this model and presented the results in Figs. \ref{F_SL_12}-\ref{F_SL_12_IC}.     In Fig. \ref{F_SL_12} we see that the model prediction for $a_0^{1\over 2}$ does not overlap with the experimental data and only is of the same order of magnitude.   The situation with $a_0^{3\over 2}$ is slightly better (Fig. \ref{F_SL_32}) where the model prediction barely touches the experimental range.   For other scattering lengths ($b_0^I$ and $c_0^I$) there are no experimental data to compare our predictions with, nevertheless we have presented them in Figs. \ref{F_SL_12} and \ref{F_SL_32}.      It is interesting to see how different contributions  to the scattering length (some of which large) conspire to result in a magnitude that is not too far from experiment.    This is shown in Fig. \ref{F_SL_12_IC} where we see that for the case of $a_0^{1\over 2}$ the large four-point contribution is balanced by other contribiutions (particularly, by $\kappa_2$ in the $s$-channel).    This is of course a known feature of the linear sigma model where such local cancellations are enforced by the underlying chiral symmetry.  

It is easy to see that the effect of the unitarizarion is negligible.    Expanding the unitarized ampltude near the threshold
\begin{eqnarray}
{\rm Re}(T_0^I)= q (\widetilde{a}_0^I+\frac{q^2}{m_\pi^2}\widetilde{b}_0^I + \frac{q^4}{m_\pi^4}\widetilde{c}_0^I+\cdots)= \frac{q(a_0^I+\frac{b_0^I}{m_\pi^2}q^2+\frac{c_0^I}{m_\pi^4}q^4+\cdots )}{1+q^2(a_0^I+\frac{b_0^I}{m_\pi^2}q^2+\frac{c_0^I}{m_\pi^4}q^4+\cdots )^2}\nonumber\\
\end{eqnarray}

Therefore
\begin{eqnarray}
\widetilde{a}_0^I&=&a_0^I,\nonumber\\
\widetilde{b}_0^I&=&b_0^I-(a_0^{I})^3 m_\pi^2,\nonumber\\
\widetilde{c}_0^I&=&c_0^I-3\, (a_0^{I})^2 b_0^I\, m_\pi^2,
\end{eqnarray}
which shows that the effects of the unitarity corrections on the scattering lengths in this framework,  are either identically 
zero or are of third order.

\clearpage
 \section{Summary and discussion}
In this work we further examined the generalized linear sigma model of Ref. \cite{global} for scalar and pseudoscalar mesons below and above 1 GeV designed to capture the global relations among all these states.        In this framework,  the underlying connections are based on mixings among two- and four-quark components which play a fundamental role in understanding the properties of scalar mesons.    The framework has been previously applied to various low-energy processes and its free parameters (in the leading order) have been determined \cite{global}.      With the same fixed parameters,  and without introducing any new ones,   the model prediction for the $I=1/2$, $J=0$, $\pi K$ scattering amplitude was calculated and then unitarized by the simple method of K-matrix.   It was shown that up to around 1 GeV the model prediction agrees with the available experimental data.   We then
examined the two poles of the unitarized scattering amplitude in this channel.   The first pole corresponds to a light and broad kappa meson with a mass around
670-770 MeV and the decay width in the range of 640-750 MeV.   This is consistent with a similar prediction in the $\pi\pi$ channel for a light and broad sigma meson, and provides further support for the generalized linear sigma model employed in this work.     The phase shift was computed and shown to have a better agreement with experiment up to slightly above 1 GeV.    With the same set of parameters,   the scattering amplitude in the $I=3/2$, $J=0$  channel was examined and shown to agree with experiment to about 1.4 GeV.   Moreover,  the $J=0$ and $I=1/2$ and $I=3/2$ scattering lengths were computed and shown that they are of the same order of magnitude as the experimental data in the $I=1/2$ channel and closer to experiment in the $I=3/2$ channel.      

The generalized linear sigma model prediction for  the kappa pole computed here is quite consistent with other findings in the literature \cite{p7,p8} (this is similar to the case of the sigma pole computed in this framework in \cite{mixing_pipi} that is consistent with the work of \cite{06_CCL}).   Overall,  the present study provides further support for the global picture of scalar and pseudoscalar mesons below and above 1 GeV \cite{global} in which it is found that the scalars below 1 GeV are closer to four-quark states and scalars above 1 GeV are closer to the conventional quark-antiquark states (and the reverse situation for the pseudoscalars).

\section*{Acknowledgments}
\vskip -.5cm
A.H.F. wishes to thank the Physics Dept. of Shiraz University for its hospitality in Summer of 2012 where this work was initiated, and is also thankful to J. Schechter for many helpful discussions.

\appendix
\section{Coupling Constants in the Single-Nonet Model}
The rotation matrices are
\begin{equation}
\left[
\begin{array}{c}
\pi^0 \\
\eta\\
\eta'
\end{array}
\right] =
R_\phi(\theta_p)
\left[
\begin{array}{c}
\phi_1^1 \\
\phi_2^2\\
\phi_3^3
\end{array}
\right]
=
\left[
\begin{array}{ccc}
{1\over \sqrt{2}} & -{1\over \sqrt{2}}      &   0\\
{a_p\over \sqrt{2}} & {a_p\over \sqrt{2}}  &  -b_p \\
{b_p\over \sqrt{2}} & {b_p\over \sqrt{2}}  &  a_p
\end{array}
\right]
\left[
\begin{array}{c}
\phi_1^1 \\
\phi_2^2\\
\phi_3^3
\end{array}
\right],
\label{Rp}
\end{equation}
with $a_p = ({{\rm cos} \theta_p - \sqrt{2} {\rm sin} \theta_p})/ {\sqrt{3}}$,
$b_p = ({\rm sin} \theta_p + \sqrt{2} {\rm cos} \theta_p) / {\sqrt{3}}$, where  $\theta_p$ is the
pseudoscalar (octet-singlet) mixing angle.  Similarly,
 \begin{equation}
\left[
\begin{array}{c}
a_0^0 \\
\sigma\\
f_0
\end{array}
\right] =
R_s(\theta_s)
\left[
\begin{array}{c}
S_1^1 \\
S_2^2\\
S_3^3
\end{array}
\right]
=
\left[
\begin{array}{ccc}
{1\over \sqrt{2}} & -{1\over \sqrt{2}}      &   0\\
{a_s\over \sqrt{2}} & {a_s\over \sqrt{2}}  &  -b_s \\
{b_s\over \sqrt{2}} & {b_s\over \sqrt{2}}  &  a_s
\end{array}
\right]
\left[
\begin{array}{c}
S_1^1 \\
S_2^2\\
S_3^3
\end{array}
\right],
\label{Rs}
\end{equation}
with $a_s = ({\rm cos} \theta_s - \sqrt{2} {\rm sin} \theta_s)/\sqrt{3}$, $b_s = ({\rm sin} \theta_s + \sqrt{2}{\rm cos} \theta_s
  / \sqrt{3}$ where $\theta_s$ is the
scalar (octet-singlet) mixing angle.
\\
The coupling constants are:
\begin{eqnarray}\label{coup}
\gamma^{(4)}_{\pi K}&=&\frac{1}{F_\pi F_K}\big[m^2_{\rm BARE}(\kappa)-m^2_K-m^2_\pi +a_s^2 m^2_{\rm BARE}(\sigma)+b_s^2 m^2_{\rm BARE}(f_0)\nonumber\\
&-&\sqrt{2}a_s b_s \big(m^2_{\rm BARE}(\sigma)-m^2_{\rm BARE}(f_0)\big)\big], \nonumber\\
 \gamma_{\kappa \pi K }&=&\frac{1}{F_K}\left( m^2_{\rm  BARE} (\kappa)- m_\pi^2 \right), \nonumber\\
\gamma_{\sigma\pi\pi} &=& \frac{1}{F_\pi}a_s \left( m^2_{\rm  BARE} (\sigma) - m_\pi^2\right), \nonumber\\
\gamma_{f_0\pi\pi} &=& \frac{1 }{F_\pi}b_s\left( m^2_{\rm BARE} (f_0) - m_\pi^2 \right), \nonumber\\
\gamma_{\sigma K K} &=& \frac{1 }{\sqrt{3}F_K}(\cos\theta_s+2\sqrt{2}\sin\theta_s)\left(  m_K^2- m^2_{\rm  BARE} (\sigma) \right), \nonumber\\
\gamma_{f_0 K K} &=& \frac{1 }{\sqrt{3}F_K}(2\sqrt{2}\cos\theta_s-\sin\theta_s)\left( m^2_{\rm BARE} (f_0)- m_K^2 \right). \nonumber\\
\label{trilinear-couplings}
\end{eqnarray}

\section{``Bare'' three- and four-point coupling constants}

\begin{eqnarray}
\left\langle \frac{\partial^3V}{\partial f_a\partial (\phi_1^3)_1\partial (\phi_3^1)_1} \right\rangle &=& 2\sqrt{2} c_4^a (2 \alpha_1-\alpha_3)
\\
\left\langle \frac{\partial^3V}{\partial f_b\partial (\phi_1^3)_1\partial (\phi_3^1)_1} \right\rangle &=& -4 c_4^a ( \alpha_1-2\alpha_3)
\\
\left\langle \frac{\partial^3V}{\partial f_b\partial (\phi_1^3)_1\partial (\phi_3^1)_1} \right\rangle &=& -4 c_4^a ( \alpha_1-2\alpha_3)
\\
\left\langle \frac{\partial^3V}{\partial f_c\partial (\phi_1^3)_1\partial (\phi_3^1)_1} \right\rangle &=& 2 \sqrt{2} e_3^a
\\
\left\langle \frac{\partial^3V}{\partial f_a\partial (\phi_1^3)_1\partial (\phi_3^1)_2} \right\rangle &=& 2 \sqrt{2} e_3^a
\\
\left\langle \frac{\partial^3V}{\partial f_a\partial (\phi_1^3)_2\partial (\phi_3^1)_1} \right\rangle &=& 2 \sqrt{2} e_3^a\\
\left\langle \frac{\partial^3V}{\partial (S_2^3)_1\partial (\phi_1^2)_1\partial (\phi_3^1)_1 } \right\rangle &=& 4\alpha_3 c_4^a
\\
\left\langle \frac{\partial^3V}{\partial (S_2^3)_2\partial (\phi_1^2)_1\partial (\phi_3^1)_1 } \right\rangle &=& -4 e_3^a
\\
\left\langle \frac{\partial^3V}{\partial (S_2^3)_1\partial (\phi_1^2)_1\partial (\phi_3^1)_2 } \right\rangle &=& -4 e_3^a \\
\left\langle \frac{\partial^3V}{\partial (S_2^3)_1\partial (\phi_1^2)_2\partial (\phi_3^1)_1 } \right\rangle &=& -4 e_3^a\\
\left\langle \frac{\partial^4V}{\partial (\phi_1^2)_1\partial (\phi_3^1)_1\partial (\phi_2^1)_1 \partial (\phi_1^3)_1} \right\rangle &=& 4 c_4^a
\end{eqnarray}

\section{Recovering current algebra limit}
In this Appendix we show how the known current algebra result for this scattering is obtained from the present model.   The four-quark fields are decoupled in the limit $d_2, e_{3}^{a}\rightarrow 0$ and $\gamma_1 \rightarrow 1$, in which:
\begin{eqnarray}
m_{\pi}^2&=&-2 c_2 +4 c_4^a \alpha_1^2 \CL 0,\nonumber \\
m_{f_1}^2&=&m_{a}^2=-2 c_ 2 +12 c_4^a \alpha_1^2,\nonumber \\
m_{f_2}^2&=&-2 c_ 2 +12 c_4^a \alpha_3^2,\nonumber \\
m_{K}^2&=&-2 c_ 2 +4 c_4^a( \alpha_1^2-\alpha_1 \alpha_3 +\alpha_3^2)\CL 0,\nonumber \\
m_{\kappa}^2&=&-2 c_ 2 +4 c_4^a( \alpha_1^2+\alpha_1 \alpha_3 +\alpha_3^2),\nonumber \\
F_{\pi}&=& 2 \alpha _{1}\nonumber, \\
m_{\eta}^2+m_{\eta^{'}}^2&=&-4 c_2-\frac{16 c_3 }{\alpha_1^2}+4 c_4^a \alpha_1^2-\frac{8 c_3 }{\alpha_3^2}+4 c_4^a \alpha_3^2
\CL -{{16 c_3}\over \alpha_1^2} - {{8 c_3}\over \alpha_3^2}.
\end{eqnarray}
Note that the above masses are expressed in terms of the  Lagrangian parameters (in  the chiral invariant part) together with  the vacuum expectation values of the scalar fields, which are in turn related to the explicit symmetry breaking term through the minimum equations,  and ensure,  for example,  that the $m_\pi$ and $m_K$ can be expressed in terms of the current quark masses and quark condensates, as expected from current algebra [and hence approach zero in the chiral limit ($C.L.$) as indicated above]. 
From the above equations we can solve for  the five model parameters:

\begin{eqnarray}\label{lpa}
\alpha_1&=&\frac{F_{\pi}}{2},\nonumber \\
\alpha_3&=&F_{\pi}\sqrt{\frac{m_{f_1}^2+ 2 m_{f_2}^2-3 m_{\pi}^2 }{12(m_{f_{1}}^2-m_{\pi}^2)}} \CL 
F_{\pi}\sqrt{\frac{m_{f_1}^2+ 2 m_{f_2}^2}{12 m_{f_{1}}^2}}
,\nonumber \\
c_2&=&\frac{1}{4}(m_{f_1}^2-3m_{\pi}^2) \CL \frac{1}{4} m_{f_1}^2,\nonumber \\
 c_3 &=& -\frac{ F_{\pi}^2(m_{f_1}^2+ 2 m_{f_2}^2-3 m_{\pi}^2)\Big(m_{f_{1}}^2-m_{f_2}^2+3 (m_{\eta}^2+m_{\eta ' }^2-2 m_{\pi}^2 )\Big) }{96(5 m_{f_{1}}^2 +4 m_{f_2}^2-9 m_{\pi}^2)}\nonumber\\
 &&
 \CL
 -\frac{ F_{\pi}^2(m_{f_1}^2+ 2 m_{f_2}^2)\Big(m_{f_{1}}^2-m_{f_2}^2+3 (-{{16 c_3}\over \alpha_1^2} - {{8 c_3}\over \alpha_3^2} )\Big) }{96(5 m_{f_{1}}^2 +4 m_{f_2}^2)},
  \nonumber \\
 c_4^a &=& \frac{m_{f_{1}}^2-m_{\pi}^2}{2 F_{\pi}^2}\CL  \frac{m_{f_{1}}^2}{2 F_{\pi}^2}.
\end{eqnarray}
Also we have:
\begin{eqnarray}
m_{\kappa}^2 &=& \frac{3}{2}(m_{f_1}^2-m_{\pi}^2)+\frac{1}{2}(-m_{f_1}^2+3 m_{\pi}^2) \CL m_{f_1}^2
\end{eqnarray}
We expect to recover the current algebra result when the scalars are decoupled as a result of becoming very heavy, i.e. in the limit $m_{f_1}=m_{f_2}=m_{a}=m_{\kappa}=m\rightarrow \infty$.    In this limit,
\begin{eqnarray}\label{lpb}
\lim _{m\rightarrow \infty} \alpha_3 &=& \frac{ F_{\pi}}{2},\nonumber \\
\lim _{m\rightarrow \infty} c_2&=&{m^2\over {4}},\nonumber \\
\lim _{m\rightarrow \infty} c_3 &=& \frac{-1}{96}F_{\pi}^2(m_{\eta}^2+m_{\eta ' }^2-2 m_{\pi}^2 ) 
\CL \frac{-1}{96}F_{\pi}^2 \left(-{{16 c_3}\over \alpha_1^2} - {{8 c_3}\over \alpha_3^2}\right),\nonumber \\
\lim _{m\rightarrow \infty} c_4^a&=&{m^2\over {2 F_\pi^2}} .\nonumber \\
\label{par_decoupling}
 \end{eqnarray}
The physical vertices (in the limit of $d_2,e_{3}^{a}\rightarrow 0$ and $\gamma_1 \rightarrow 1$) become:
\begin{eqnarray}\label{gf2pipi}
\gamma^{(4)}&=&4 c_4^a, \nonumber \\
\gamma_{f_1 \pi \pi }&=&4 c_4^a \alpha_1,\nonumber \\
\gamma_{f_2 \pi \pi }&=&0,\nonumber \\
\gamma_{f_1 K K }&=&4 c_4^a (2\alpha_1-\alpha_3),\nonumber \\
\gamma_{f_2 K K}&=&-4\sqrt{2} c_4^a ( \alpha_1-2\alpha_3),\nonumber \\
\gamma_{\kappa \pi K }&=&4 c_4^a \alpha_3 .\nonumber \\
\end{eqnarray}
which together with (\ref{par_decoupling}),

\begin{eqnarray}\label{gf2pipi}
\gamma^{(4)}&=&\frac{2(m^2-m_{\pi}^2)}{ F_{\pi}^2} \CL \frac{2 m^2}{ F_{\pi}^2} , \nonumber \\
\gamma_{f_1 \pi \pi }&=&\frac{m^2-m_{\pi}^2}{ F_{\pi}^2} \CL \frac{m^2}{ F_{\pi}^2} ,\nonumber \\
\gamma_{f_2 \pi \pi }&=&0,\nonumber \\
\gamma_{f_1 K K }&=&\frac{m^2-m_{\pi}^2}{ F_{\pi}^2}\CL \frac{m^2}{ F_{\pi}^2},\nonumber \\
\gamma_{f_2 K K}&=&\frac{\sqrt{2}(m^2-m_{\pi}^2)}{ F_{\pi}^2} \CL \frac{\sqrt{2} m^2}{ F_{\pi}^2},\nonumber \\
\gamma_{\kappa \pi K }&=&\frac{m^2-m_{\pi}^2}{F_{\pi}^2} \CL \frac{m^2}{F_{\pi}^2}.\nonumber \\
\end{eqnarray}

Each individual  decay amplitude inherits the scalar mass dependency via the physical vertices and propagators.  The four-point amplitude will have the scalar mass dependency

\begin{equation}
M_{4p}=\xi_0+\xi_1 m^2.
\end{equation}
The isosinglet scalar contribution  has the general structure
\begin{equation}
M_{f_j}=\gamma_{f_j\pi\pi}\gamma_{f_j K K}\times (\rm{propagator}),
\end{equation}
with
\begin{eqnarray}
\gamma_{f_j\pi\pi}\gamma_{f_j K K}&=&\rho_0+\rho_1 m^2+\rho_2 m^4,\nonumber\\
\rm{propagator}&=&\frac{1}{m^2+z}\simeq \frac{1}{m^2}-\frac{z}{m^4}+ {\cal O} (\frac{1}{m^6}).
\end{eqnarray}
Thus
\begin{equation}
\lim _{m\rightarrow \infty} M_{f_j}=\rho_1 - z \rho_2 +\rho_2 m^2.
\end{equation}
Similarly for the $\kappa$ contribution
\begin{equation}
M_{\kappa}=\gamma_{\kappa\pi K}^2\Big[\frac{3}{2}\frac{1}{m^2+y_1}-\frac{1}{2}\frac{1}{m^2+y_2}\Big],
\end{equation}
with
\begin{eqnarray}
\gamma_{\kappa\pi K}^2=\delta_0 +\delta_1 m^2 +\delta_2 m^4, \nonumber\\
\frac{1}{m^2+y_i}
\simeq\frac{1}{m^2}-\frac{y_i}{m^4}+{\cal O} (\frac{1}{m^6}).
\end{eqnarray}
Thus
\begin{equation}
\lim _{m\rightarrow \infty} M_{\kappa}= \delta_1-(\frac{3}{2} y_1-\frac{1}{2} y_2)\delta_2 + \delta_2 m^2.
\end{equation}
Now putting everything together, we expect:
\begin{equation}
\lim_{m\rightarrow\infty} M_{\rm{total}}=M_{\rm{C.A.}}
\end{equation}
which implies that the following two sum rules must be upheld
\begin{eqnarray}
\xi_0+\rho_1-z \rho_2+\delta_1-(\frac{3}{2} y_1-\frac{1}{2} y_2)\delta_2&=&M_{\rm{C.A.}},\nonumber\\
\xi_1 +\rho_2 +\delta_2 &=&0.
\end{eqnarray}
We find that the second sum-rule is identically upheld, and the first one gives:
\begin{equation}
M_{\rm{C.A.}}=\frac{-4 m_{\pi}^2(3\cos{\theta}+1)+(3\cos{\theta}+5)s}{4 F_{\pi}^2} \CL 
\frac{(3\cos{\theta}+5)s}{4 F_{\pi}^2}.
\end{equation}
in agreement with Eq. (3.2) of Ref. \cite{BFSS1}.


\end{document}